\documentclass[twocolumn]{aastex631}
\usepackage{amsmath}
\usepackage{multirow}
\usepackage{sidecap}

\begin{document}

\title{NIRSpec View of the Appearance and Evolution of Balmer Breaks and the Transition from Bursty to Smooth Star Formation Histories from Deep Within the Epoch of Reionization to Cosmic Noon}

\correspondingauthor{Danial Langeroodi}
\email{danial.langeroodi@nbi.ku.dk}

\author[0000-0001-5710-8395]{Danial Langeroodi}
\affil{DARK, Niels Bohr Institute, University of Copenhagen, Jagtvej 155, 2200 Copenhagen, Denmark}

\author[0000-0002-4571-2306]{Jens Hjorth}
\affil{DARK, Niels Bohr Institute, University of Copenhagen, Jagtvej 155, 2200 Copenhagen, Denmark}

\begin{abstract}
Theoretical models and observational evidence suggest that high-redshift galaxies grow under the bursty mode of star formation, with large temporal star formation rate (SFR) fluctuations around some mean value. From an observational perspective, it has not been clear at which redshift and stellar population characteristics the transition from bursty to smooth star formation occurs. Here, we investigate these using a uniformly reduced sample of NIRSpec prism spectra of 631 galaxies at $3 < z_{\rm spec} < 14$, stacked in 8 redshift and 8 UV slope bins. We evaluate the burstiness of star formation histories using the Balmer break strengths as well as the ratios of SFRs as measured from the emission lines to those measured from the UV continua. The break strength increases monotonically from $z = 10$ to $z = 3$, and from $\beta_{\rm UV} = -3.0$ to $\beta_{\rm UV} = 0.0$. The break strength is tightly anti-correlated with specific SFR (sSFR), and in dusty galaxies, strongly correlated with dust attenuation. Based on the SFR ratios, we find that bursty star formation thrives in the highest redshift, bluest, and lowest stellar mass galaxies, which exhibit the highest sSFRs. The burstiness appears to plateau at $z > 6$, suggesting that we might be observing the peak of star formation burstiness at these redshifts. The $z < 4$ galaxies do not appear particularly bursty, suggesting that the smooth mode of star formation starts taking over right before cosmic noon. As galaxies mature and develop redder UV colors and more pronounced Balmer breaks, their ability to sustain star formation over longer timescales increases, signalling their transition from bursty to smooth star formation. 
\end{abstract}


\section{Introduction} \label{sec: intro}

While the process of star formation in galaxies is relatively well understood, the mechanisms that regulate it and drive their transition from blue and star-forming to red and passive are under active debate. The proposed contributors include gas removal and heating by stellar and AGN feedback \citep[for stellar feedback, see e.g.,][and for AGN feedback, see e.g., \citealp{2005Natur.433..604D, 2006MNRAS.370..645B, 2008MNRAS.391..481S, 2012ARA&A..50..455F, 2012MNRAS.425..605F, 2017MNRAS.464.2840A, 2018MNRAS.475.1288S, 2018MNRAS.479.4056W, 2020MNRAS.493.1888T, 2020MNRAS.499..768Z, 2022MNRAS.512.1052P}]{2009ApJ...695..292C, 2012MNRAS.421.3522H, 2014MNRAS.445..581H, 2015MNRAS.451.2757W, 2015NewAR..68....1D, 2015MNRAS.450..504M, 2017MNRAS.466.1093G, 2018MNRAS.475.3511G, 2020MNRAS.494.3971M, 2022ApJ...936..137O, 2024arXiv240319843S}; other mass-dependent mechanisms, such as virial shock heating of the infalling gas \citep{2006MNRAS.368....2D, 2006MNRAS.370.1651C, 2007MNRAS.380..339B, 2013ApJ...762L..31B, 2015MNRAS.447..374G}; increased velocity dispersion and structural transformations that stabilize the gas against fragmentation \citep{2009ApJ...707..250M, 2010MNRAS.404.2151C, 2013MNRAS.432.1914M, 2019ApJ...872...50L, 2023ApJ...953...27L, 2024ApJ...964..129M}; and the processes associated with dense environments, such as ram pressure, tidal interactions in clusters, thermal evaporation, starvation, and interactions with other galaxies \citep{2006PASP..118..517B, 2010ApJ...721..193P, 2014MNRAS.438..717K, 2016ApJ...825..113D, 2017ApJ...844...48P, 2018ApJ...857...71B, 2018MNRAS.478..548S, 2019ApJ...876...40P, 2021MNRAS.500.4004D, 2024ApJ...961...39S}.

Despite recent advances, the relative importance of different mechanisms in regulating star formation and its eventual cessation remains unclear. Ambiguities include the relative contribution of galaxy internal versus environmental processes \citep{2010ApJ...721..193P, 2015ARA&A..53...51S,  2018MNRAS.475..648P, 2021MNRAS.500.4004D}; efficiency and scale of AGN feedback as well as the type of feedback (e.g., thermal, radiative, kinetic, mechanical) it induces \citep[e.g., different implementations of AGN feedback in][and \citealp{2019MNRAS.486.2827D}]{2012MNRAS.420.2662D, 2014ApJ...789..150G, 2014MNRAS.445..581H, 2015MNRAS.446..521S, 2017MNRAS.464.2840A, 2017MNRAS.465.3291W}; the stellar feedback prescription, namely, the specific roles of protostellar jets, radiation pressure, thermal heating, stellar winds, and supernovae \citep{2013ApJ...770...25A, 2013MNRAS.428..129S, 2018MNRAS.480..800H, 2019MNRAS.485.3317S, 2024arXiv240319843S}; and the non-linear combination of feedback mechanisms \citep{2014MNRAS.442.3013K, 2014MNRAS.445..581H, 2018MNRAS.481.3325F}. 

Calibrating and comparing the theoretical models and simulations against a diverse array of observations is key to breaking these degeneracies \citep{2015MNRAS.450.1937C, 2017MNRAS.465.2936M, 2023ARA&A..61..473C}. Different star formation regulation and quenching mechanisms operate over significantly different temporal scales, often separated by several orders of magnitude. Therefore, constraints from high-redshift observations are crucial in informing the theoretical models. In this regime, the limited lifetimes of galaxies simplify the complexities of star formation regulation and quenching by imposing stringent limits on the timescales of mechanisms that can contribute. The most rapid mechanisms, such as stellar and potentially AGN feedback, are expected to play pivotal roles in setting the stellar mass assembly and morphological fates of galaxies emerging in the early Universe \citep[see, e.g.,][]{2023MNRAS.525.5388B, 2023MNRAS.523.3201D, 2023MNRAS.525.2241H, 2024arXiv240208717S, 2024arXiv240402815W, 2024arXiv240402537C}.

High-redshift galaxies, as well as low-mass galaxies, are predicted to exhibit ``bursty'' star formation histories characterized by frequent transitions between periods of elevated and suppressed star formation \citep{2015MNRAS.454.2691M, 2018MNRAS.480.4842C, 2018MNRAS.473.3717F, 2020MNRAS.497..698T, 2021MNRAS.501.4812F, 2023A&A...677L...4P, 2023MNRAS.525.2241H, 2023arXiv231016895B, 2020MNRAS.498..430I, 2024ApJ...961...53I}. This prediction is supported by growing empirical evidence \citep{2016ApJ...833..254S, 2017MNRAS.466...88S, 2019ApJ...881...71E, 2019ApJ...884..133F, 2021MNRAS.500.5229E, 2023MNRAS.524.2312E, 2023arXiv230605295E, 2022ApJ...930..128R, 2023ApJ...952..143R, 2023arXiv230602470L, 2023arXiv231112691C, EVOLFMR, 2024MNRAS.52711372A, 2024ApJ...964..150D, 2024MNRAS.529.4728D}. Observational constraints on the duration, frequency, and amplitudes of star formation burst fluctuations, along with the redshifts, environments, stellar masses, and morphological types where the bursty mode of star formation thrives can inform the timescales of gas infall, cooling, and feedback, and their interplay with the galaxy's gravitational potential \citep[see][and the references above]{2023MNRAS.525.2241H}. 

The ``burstiness'' of star formation can be evaluated through several observables. For instance, this can be done by comparing the star formation rates (SFRs) as measured from the Balmer emission lines to those measured from the UV continuum. Since the former traces the star formation over the past $\sim 10$ Myr and the latter over the past $\sim 100$ Myr, this evaluates if galaxies undergo regular bursts of star formation while constraining the amplitude of these bursts. Similarly, the strengths of Balmer breaks provide a measure of burstiness. Detecting a pronounced Balmer break indicates that recent star formation is not high enough for the young massive stars to outshine the more mature stellar population of the galaxies. This is characteristic of galaxies with low specific SFRs (sSFR; SFR/M$_{\star}$). When accompanied by suppressed Balmer lines, a strong Balmer break is interpreted as a quenched galaxy. As a consequence of their bursty star formation histories, the observed passive galaxies at $z > 5$ are most likely experiencing temporarily quiescent periods \citep{2024MNRAS.527.2139D, 2023ApJ...954L..11G, 2024ApJ...964...76G}. The number densities and star formation histories of these galaxies constrain the duration and frequency of the bursts. 

Detecting and characterizing the high-redshift low-sSFR/quiescent galaxies with NIRCam photometry can be challenging. Depending on the source redshift and photometric coverage, the Balmer break strength can be degenerate with the equivalent widths of strong adjacent emission lines such as H$\beta$ and the [O\,\textsc{iii}]4959,5007\AA\ doublet. This is exacerbated by photometric redshift estimation inaccuracies. Several methods can be employed to break this degeneracy, including a careful selection of photometric redshifts \citep{2024MNRAS.52711627T}; addition of NIRCam medium-band imaging, which isolates the emission lines from the continuum \citep{2023arXiv231012340E}; and addition of deep MIRI photometry, which probes the continuum elevation on the red side of H$\beta$ and [O\,\textsc{iii}] \citep[][Langeroodi et al.\ in prep.; Östlin et al.\ in prep.]{2023arXiv231212207A, 2024arXiv240302399W}. Nonetheless, NIRSpec follow-ups are often our only means for confirming the quiescent candidates. Other burstiness diagnostics such as comparing emission line SFRs with those from the rest-UV continuum are not accessible without NIRSpec spectroscopy. 

Through emission-line diagnostics, NIRSpec multi-object spectroscopy \citep[MOS;][]{2022A&A...661A..80J, 2022A&A...661A..81F} has proved particularly efficient at probing the interstellar medium conditions of galaxies out to $z \sim 10$ \citep{2022A&A...665L...4S, 2023MNRAS.518..425C, 2023arXiv230408516C, 2023MNRAS.525.2087B, 2023Sci...380..416W, 2023ApJ...957...39L, 2023NatAs...7.1517H, 2023ApJS..269...33N, EVOLFMR, 2023A&A...677A.115C, 2023ApJ...950L...1S, 2024arXiv240116934B}. In addition, the prism disperser has provided a unique view of high-$z$ galaxy compositions through high S/N rest-UV and rest-optical continuum detections. Thus far, the latter has led to confirmation of the V-shaped continua of little red dots \citep{2023arXiv230805735F, 2023arXiv231203065K, 2024ApJ...964...39G, 2024arXiv240302304W} with rest-optical continuum turnovers driven by their obscured AGNs \citep{2023ApJ...954L...4K, 2024ApJ...964...39G, 2023arXiv230605448M, 2023arXiv231203065K, 2023arXiv230607320L, 2024ApJ...964...39G, 2024arXiv240108782P, 2024arXiv240403576K}; discovery of the 2175\AA\ attenuation bump feature at redshifts as high as $z \sim 6.7$ \citep{2023Natur.621..267W}; and discovery of surprisingly evolved and passive galaxies, given their redshifts, through detection of strong Balmer breaks \citep{2023arXiv230805606G, 2023Natur.619..716C, 2023arXiv230214155L, 2023ApJ...949L..23S, 2024arXiv240405683D}.

In this work, we leverage the capabilities of the NIRSpec prism multi-object spectroscopy to investigate the appearance and redshift evolution of Balmer breaks, as well as the burstiness of star formation histories for $3 < z_{\rm spec} < 10$ galaxies. For this purpose, we uniformly reduce and calibrate the publicly available NIRSpec prism data to compile a sample of 624 galaxies at $z_{\rm spec} > 3$ (Section \ref{sub: nirspec and nircam}). We stack the obtained spectra in 8 redshift and 8 UV slope bins (Section \ref{sub: stacking}). We use the stacks to probe the redshift and UV slope trends of the Balmer break strength (Section \ref{sec: Balmer breaks}). Through spectral energy distribution fitting (Section \ref{sub: SED}), we investigate the correlation between Balmer break strength and stellar population parameters such as specific SFR or dust attenuation (Section \ref{sub: sSFR and Av}). Moreover, we compare the specific SFRs as measured from Balmer lines with those measured from the rest-UV continuum to evaluate the burstiness of star formation histories of our stacks (Section \ref{sec: burstiness}). We discuss and summarize our findings in Sections \ref{sec: discussion} and \ref{sec: conclusion}. 

\begin{deluxetable*}{lccccccc}
\tablewidth{0pt}
\tablecaption{Overview of the Redshift Stacks of NIRSpec Prism Spectra.}
\label{table: redshift stacks}
\tablehead{
\colhead{Name} &
\colhead{\#} &
\colhead{Redshift Range (Median)} &
\colhead{sSFR$_{\rm lines}$} & 
\colhead{sSFR$_{\rm SED}$} &
\colhead{$A_{\rm V}$} &
\colhead{Strength of Balmer Break} \\ 
\colhead{} & 
\colhead{} & 
\colhead{} &
\colhead{Gyr$^{-1}$} &
\colhead{Gyr$^{-1}$} &
\colhead{mag} &
\colhead{$f_{\nu}(4225{\rm\AA})/f_{\nu}(3565{\rm\AA})$}
}
\startdata
$z[3.0,3.5]$  & 125 & 3.0 to 3.5 (3.28)  & $0.67^{+0.10}_{-0.10}$ & $0.67^{+0.08}_{-0.08}$ & $0.65^{+0.04}_{-0.05}$ & $1.53 \pm 0.13$ \\
$z[3.5,4.0]$  & 81  & 3.5 to 4.0 (3.72)  & $0.89^{+0.06}_{-0.06}$ & $0.71^{+0.10}_{-0.12}$ & $0.36^{+0.04}_{-0.04}$ & $1.39 \pm 0.09$ \\
$z[4.0,4.5]$  & 83  & 4.0 to 4.5 (4.29)  & $1.10^{+0.08}_{-0.09}$ & $0.79^{+0.08}_{-0.07}$ & $0.23^{+0.04}_{-0.04}$ & $1.30 \pm 0.09$ \\
$z[4.5,5.0]$  & 73  & 4.5 to 5.0 (4.74)  & $1.29^{+0.10}_{-0.09}$ & $0.79^{+0.16}_{-0.17}$ & $0.27^{+0.08}_{-0.08}$ & $1.16 \pm 0.08$ \\
$z[5.0,5.5]$  & 68  & 5.0 to 5.5 (5.17)  & $1.44^{+0.07}_{-0.12}$ & $0.89^{+0.07}_{-0.14}$ & $0.21^{+0.05}_{-0.03}$ & $1.22 \pm 0.11$ \\
$z[5.5,6.0]$  & 60  & 5.5 to 6.0 (5.76)  & $1.44^{+0.11}_{-0.15}$ & $0.96^{+0.07}_{-0.15}$ & $0.11^{+0.05}_{-0.05}$ & $1.09 \pm 0.11$ \\
$z[6.0,7.0]$  & 72  & 6.0 to 7.0 (6.36)  & $1.91^{+0.11}_{-0.12}$ & $1.39^{+0.09}_{-0.08}$ & $0.11^{+0.04}_{-0.03}$ & $0.94 \pm 0.12$ \\
$z[7.0,10.0]$ & 62  & 7.0 to 10.0 (7.82) & $2.16^{+0.21}_{-0.19}$ & $1.55^{+0.19}_{-0.20}$ & $0.12^{+0.07}_{-0.07}$ & $0.69 \pm 0.13$ \\
\enddata
\end{deluxetable*}

\begin{deluxetable*}{lccccccc}
\tablewidth{0pt}
\tablecaption{Overview of the UV Slope Stacks of NIRSpec Prism Spectra.}
\label{table: beta stacks}
\tablehead{
\colhead{Name} &
\colhead{\#} &
\colhead{UV Slope Range} &
\colhead{sSFR$_{\rm lines}$} & 
\colhead{sSFR$_{\rm SED}$} &
\colhead{$A_{\rm V}$} &
\colhead{Strength of Balmer Break} \\ 
\colhead{} & 
\colhead{} & 
\colhead{(Median)} &
\colhead{Gyr$^{-1}$} &
\colhead{Gyr$^{-1}$} &
\colhead{mag} &
\colhead{$f_{\nu}(4225{\rm\AA})/f_{\nu}(3565{\rm\AA})$}
}
\startdata
$\beta_{\rm UV}[-3.00,-2.50]$  & 63  & $-3.00$ to $-2.50$ ($-2.68$)  & $2.16^{+0.29}_{-0.23}$ & $1.50^{+0.26}_{-0.26}$ & $0.00^{+0.01}_{-0.00}$ & $0.80 \pm 0.13$ \\
$\beta_{\rm UV}[-2.50,-2.25]$  & 65  & $-2.50$ to $-2.25$ ($-2.35$)  & $1.77^{+0.08}_{-0.08}$ & $1.07^{+0.11}_{-0.10}$ & $0.00^{+0.01}_{-0.00}$ & $0.95 \pm 0.07$ \\
$\beta_{\rm UV}[-2.25,-2.00]$  & 107 & $-2.25$ to $-2.00$ ($-2.13$)  & $1.25^{+0.10}_{-0.11}$ & $0.85^{+0.12}_{-0.13}$ & $0.03^{+0.04}_{-0.01}$ & $1.09 \pm 0.07$ \\
$\beta_{\rm UV}[-2.00,-1.75]$  & 103 & $-2.00$ to $-1.75$ ($-1.87$)  & $1.47^{+0.09}_{-0.12}$ & $0.90^{+0.12}_{-0.13}$ & $0.25^{+0.05}_{-0.05}$ & $1.06 \pm 0.07$ \\
$\beta_{\rm UV}[-1.75,-1.50]$  & 100 & $-1.75$ to $-1.50$ ($-1.65$)  & $1.13^{+0.12}_{-0.08}$ & $0.73^{+0.09}_{-0.08}$ & $0.37^{+0.07}_{-0.03}$ & $1.22 \pm 0.07$ \\
$\beta_{\rm UV}[-1.50,-1.25]$  & 62  & $-1.50$ to $-1.25$ ($-1.40$)  & $1.06^{+0.07}_{-0.07}$ & $0.71^{+0.06}_{-0.05}$ & $0.52^{+0.03}_{-0.04}$ & $1.32 \pm 0.07$ \\
$\beta_{\rm UV}[-1.25,-1.00]$  & 47  & $-1.25$ to $-1.00$ ($-1.14$)  & $0.87^{+0.12}_{-0.12}$ & $0.64^{+0.11}_{-0.09}$ & $0.67^{+0.04}_{-0.04}$ & $1.38 \pm 0.08$ \\
$\beta_{\rm UV}[-1.00,0.00]$   & 51  & $-1.00$ to $0.00$  ($-0.67$)  & $0.77^{+0.15}_{-0.16}$ & $0.62^{+0.13}_{-0.15}$ & $0.99^{+0.08}_{-0.09}$ & $1.56 \pm 0.10$ \\
\enddata
\end{deluxetable*}

\section{Data} \label{sec: data}

\subsection{NIRSpec \& NIRCam data} \label{sub: nirspec and nircam}

In this work, we compile a sample of 631 NIRSpec prism spectra at $3 < z_{\rm spec} < 14$ from the public cycle 1 and 2 JWST data. All the spectra are uniformly reduced through the pipeline described in \cite{EVOLFMR} and visually inspected for accurate spectroscopic redshifts. An overview of the included JWST programs and the relevant references are presented in Table 1 of Langeroodi et al.\ (in prep.). This sample is mostly identical to that of Langeroodi et al.\ (in prep.) but with the requirement for high-S/N UV continua relaxed, resulting in a larger sample size. 

The NIRSpec prism spectra of individual galaxies were reduced in \cite{EVOLFMR} and Langeroodi et al.\ (in prep.), where the details of our reduction and flux calibration pipelines are presented respectively. Here, we provide a brief overview of these pipelines. The stage-one reduction (raw data to count-rate images) was achieved using the STScI JWST pipeline \citep{2022A&A...661A..81F, bushouse_howard_2023_7795697}. The stage-2 and -3 reduction as well as the optimal extraction of 1D spectra were carried out using the \texttt{msaexp} pipeline \citep{msaexp}. Before visual inspection, the spectroscopic redshifts were fitted using \texttt{EAZY} \citep{eazy}, first by searching for the Ly$\alpha$ break and then by fitting the emission lines. 

We corrected the residual slit-loss by scaling the optimally extracted 1D spectra to match the publicly available NIRCam photometry \citep{EVOLFMR}. In brief, we adopted the aperture-corrected PSF-matched (to the F444W imaging) photometry from \cite{EVOLFMR} in $0.3\arcsec$ apertures. To calibrate each 1D spectrum, we re-scaled it with a linear calibration function (as a function of wavelength) inferred as the difference between NIRCam photometry and the projection of 1D spectrum onto NIRCam filters; this procedure is repeated until either the 1D spectrum is calibrated within $1\sigma$ of the NIRCam photometry or a maximum of 5 iterations is reached. Next, we assumed a 2nd order polynomial calibration function and repeated the same procedure. 

Since in this work we are interested in studying the strength of Balmer breaks for $z > 3$ galaxies, the known broad-line AGNs and Little Red Dots \citep{2023ApJ...959...39H, 2023ApJ...954L...4K, 2023arXiv230801230M, 2023arXiv230805735F, 2023ApJ...957L..27L, 2023ApJ...957L...7K, 2023arXiv231203065K, 2024ApJ...963..128B, 2024ApJ...964...39G} are excluded from our sample. This is because at wavelengths redder than rest-4000\AA\ the spectra of these sources tend to be elevated due to the contribution by an obscured AGN \citep{2023ApJ...954L...4K, 2024arXiv240403576K, 2023arXiv230605448M, 2023arXiv231203065K, 2023arXiv230607320L, 2024ApJ...964...39G, 2024arXiv240108782P}; this elevation can systematically bias the Balmer break measurements high. 

\subsection{Stacking} \label{sub: stacking}

\begin{figure*}
    \centering
    \includegraphics[width=16.8cm]{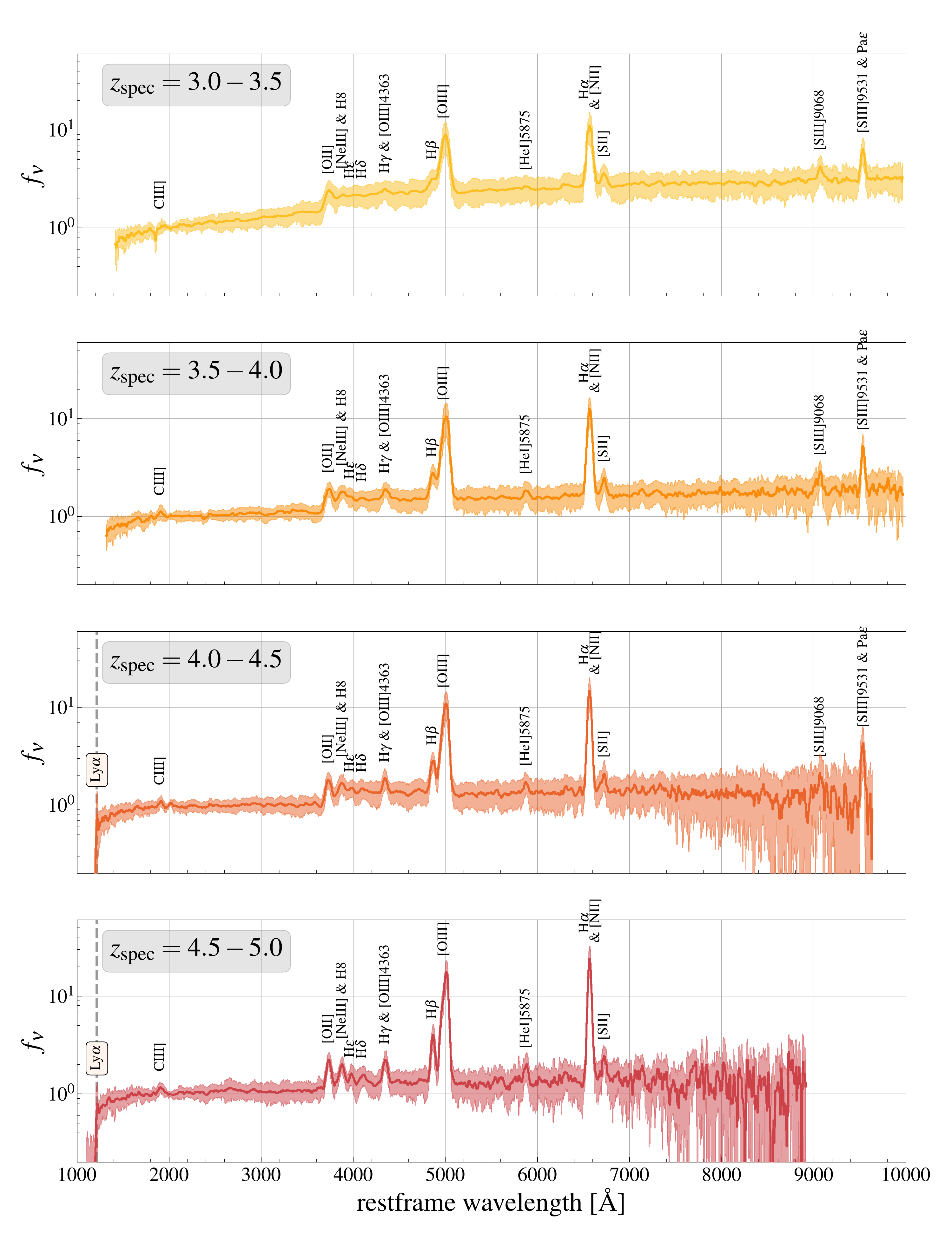}
    \caption{Stacked NIRSpec prism spectra in 4 redshift bins ($3<z<5$). The redshift range of each stack is indicated at the upper left of its corresponding panel. Where covered by the stacks, prominent emission lines are indicated by their names.}
    \label{fig: redshift stacks 1}
\end{figure*}

\begin{figure*}
    \centering
    \includegraphics[width=16.8cm]{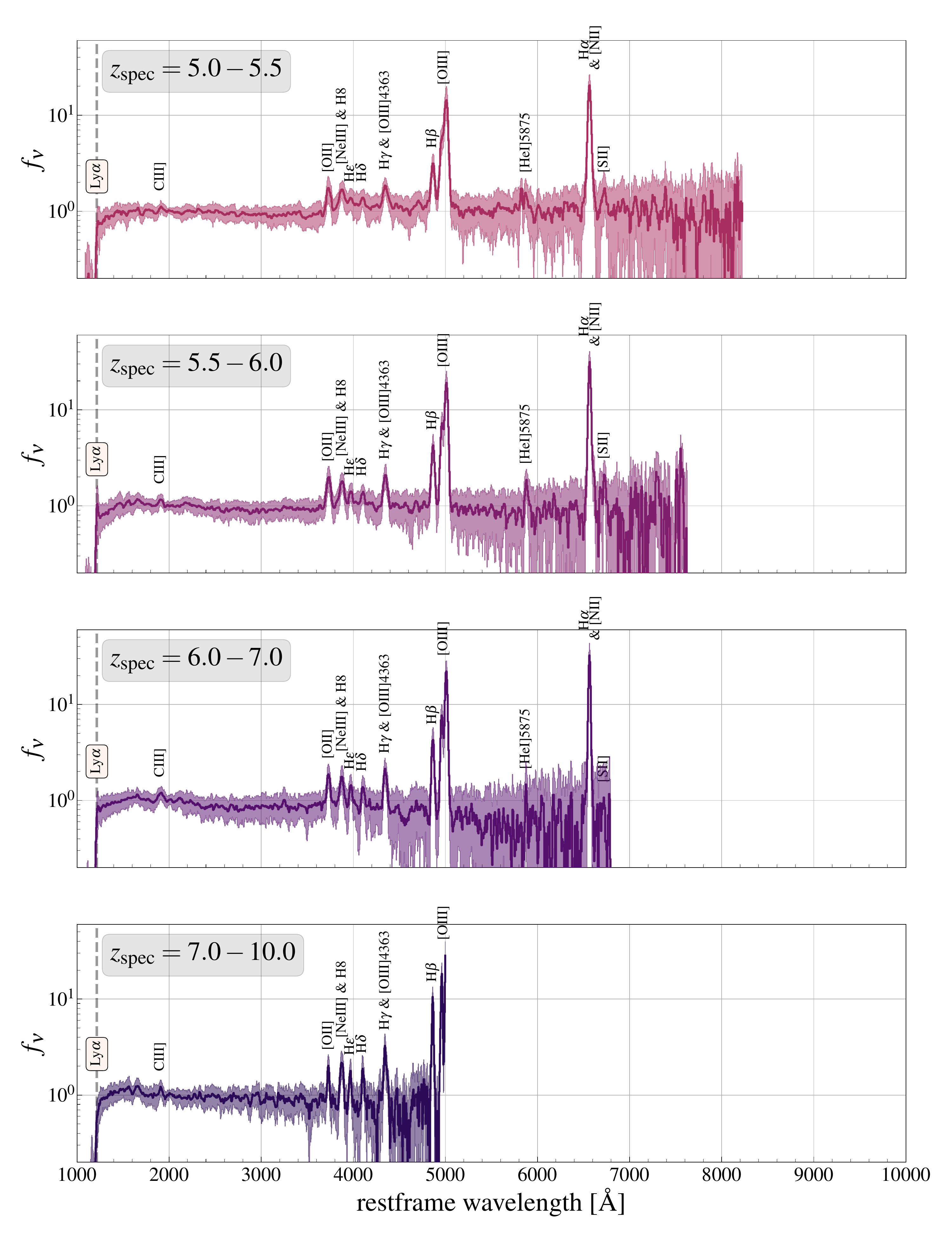}
    \caption{Stacked NIRSpec prism spectra in 4 redshift bins ($5<z<10$). The redshift range of each stack is indicated at the upper left of its corresponding panel. Where covered by the stacks, prominent emission lines are indicated by their names.}
    \label{fig: redshift stacks 2}
\end{figure*}

\begin{figure*}
    \centering
    \includegraphics[width=16.8cm]{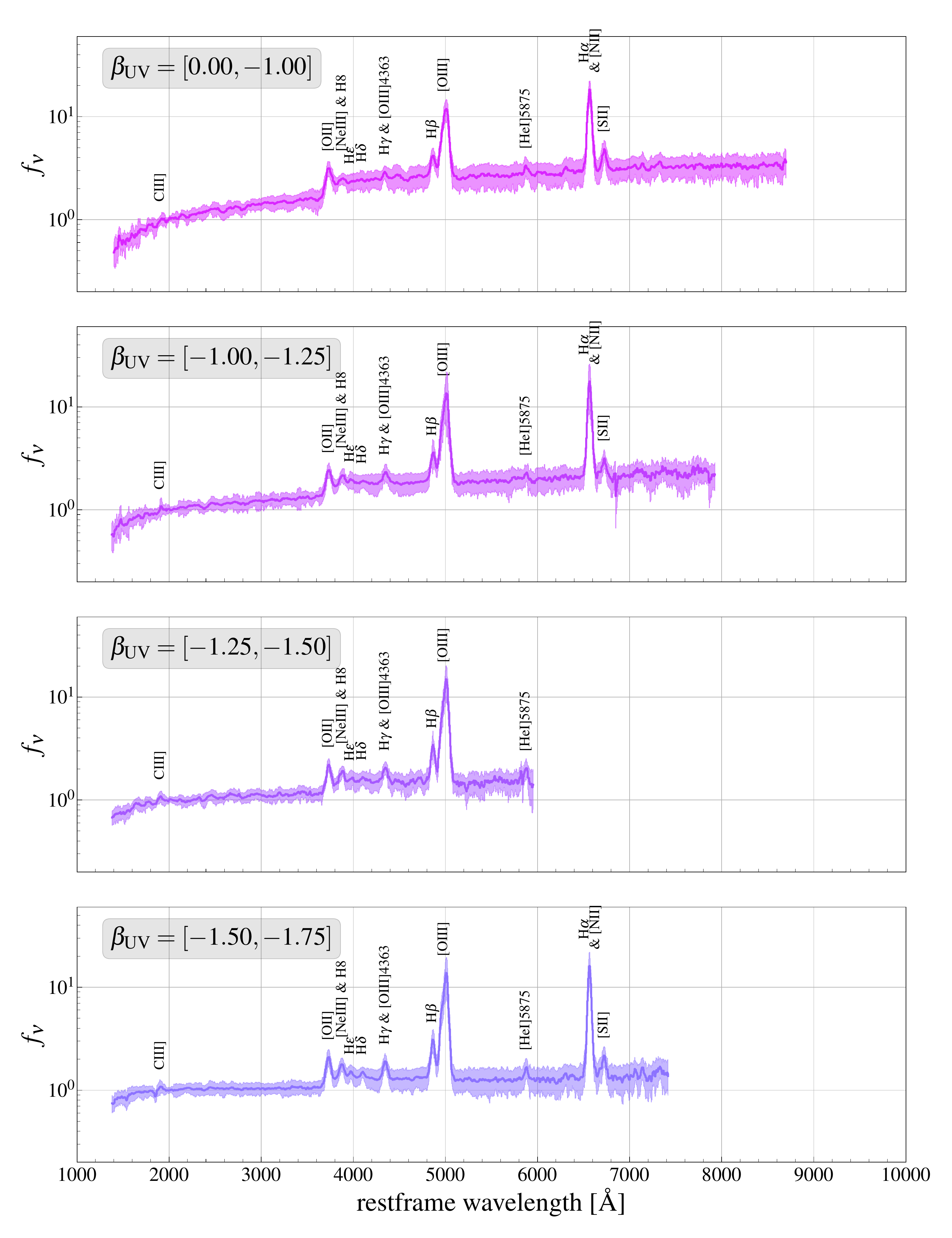}
    \caption{Stacked NIRSpec prism spectra in 4 UV slope bins ($-1.75<\beta_{\rm UV}<0$). The UV slope range of each stack is indicated at the upper left of its panel. Where covered by the stacks, prominent emission lines are indicated by their names.}
    \label{fig: beta stacks 1}
\end{figure*}

\begin{figure*}
    \centering
    \includegraphics[width=16.8cm]{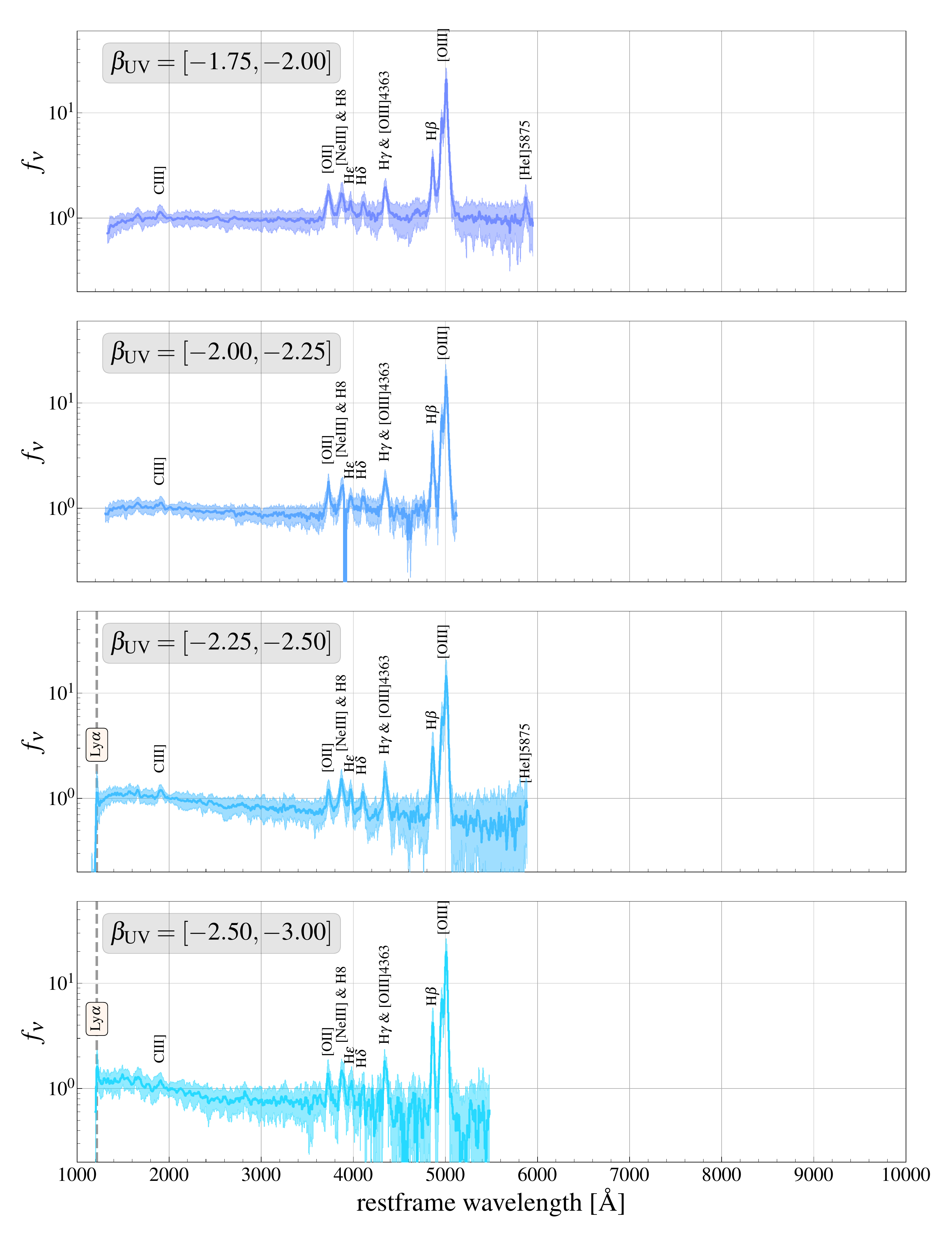}
    \caption{Stacked NIRSpec prism spectra in 4 UV slope bins ($-3<\beta_{\rm UV}<-1.75$). The UV slope range of each stack is indicated at the upper left of its panel. Where covered by the stacks, prominent emission lines are indicated by their names.}
    \label{fig: beta stacks 2}
\end{figure*}

We prepared two sets of stacked spectra, each consisting of 8 stacks; one set consists of stacks in redshift bins and the other of stacks in UV slope bins. An overview of the stacks is provided in Tables \ref{table: redshift stacks} and \ref{table: beta stacks}. The stacks are shown in Figures \ref{fig: redshift stacks 1}, \ref{fig: redshift stacks 2}, \ref{fig: beta stacks 1}, and \ref{fig: beta stacks 2}. Within each set, the bins are selected under two criteria to ensure that i) each bin comprises a minimum of 50 galaxies, and ii) the bins cover equally-spaced intervals in the binning parameter ($z_{\rm spec}$ and $\beta_{\rm UV}$). The latter condition is relaxed at $ z > 6$ as well as both ends of our UV slope coverage, where larger bins are selected to enable stacking at least 50 galaxies. The former condition is relaxed only for one UV slope bin that contains 47 galaxies (the $-1.25 < \beta_{\rm UV} < -1.00$ bin). We adopt the UV slope measurements from Langeroodi et al.\ (in prep.), where $\beta_{\rm UV}$ and $M_{\rm UV}$ are measured from the NIRSpec prism spectra of a subset of our sample consisting of 598 galaxies with high S/N UV continua.

To generate each bin's stack, we rest-framed its spectra, resampled them to a common wavelength grid, normalized them, and weighed them by their variance before coaddition. The spectral sampling of the common grid is chosen to match the highest rest-FWHM of the spectra in that bin. We used the \texttt{SpectRes} \citep{2017arXiv170505165C} package to resample the spectra and their uncertainties to the common grid. Each spectrum is scaled to a flux density of 1 at rest-2000\AA. The spectra are then weighed by their variance at rest-2000\AA\ and coadded to generate the stack. The $1\sigma$ uncertainty of each spectral pixel is determined by the variance-weighted $1\sigma$ distribution of the flux densities in that pixel. 

\begin{figure*}
    \centering
    \includegraphics[width=13.5cm]{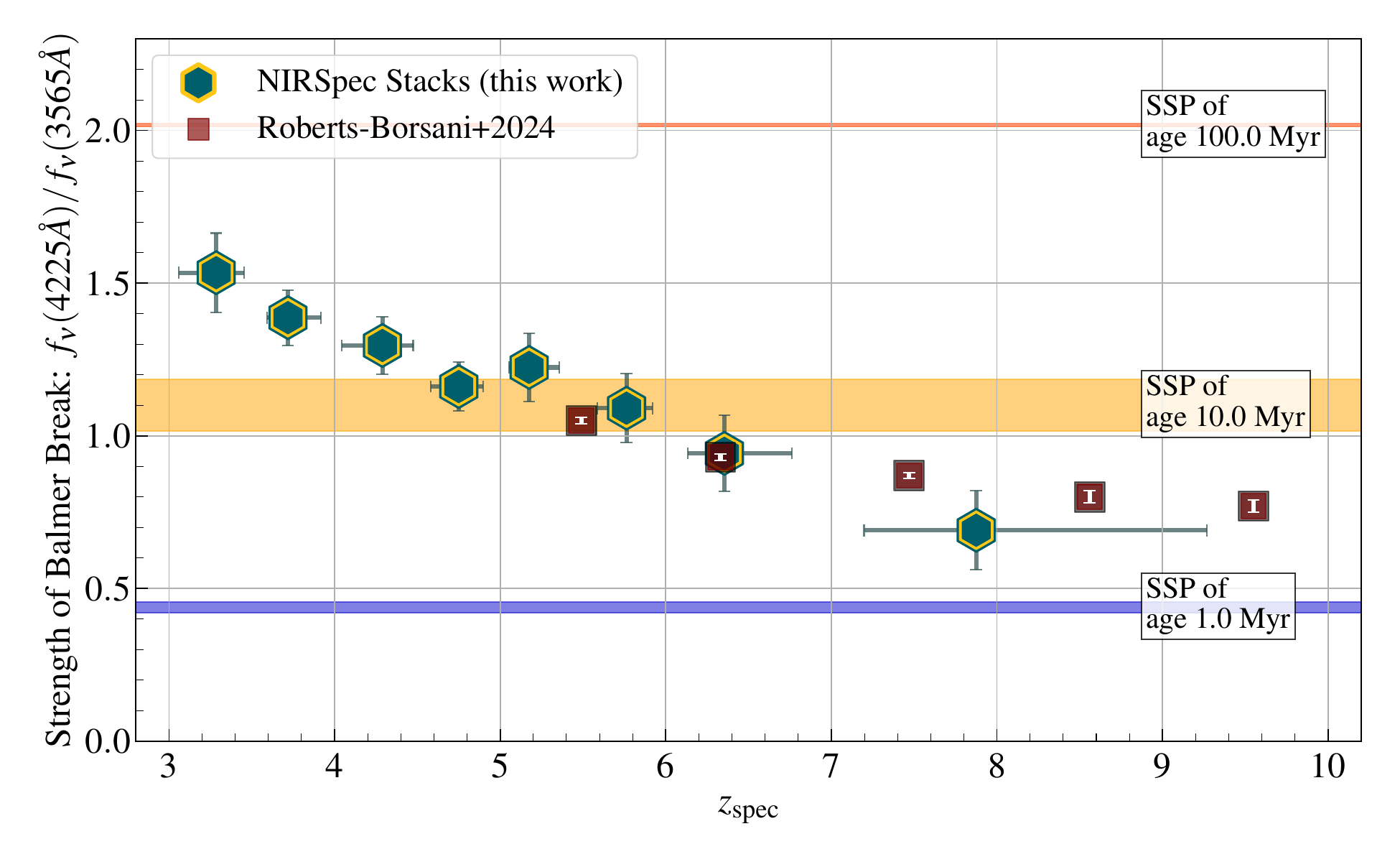}
    \caption{Evolution of the Balmer break strength from $z = 10$ to $z = 3$. The dark-green diamonds show the break strengths as measured from our stacked spectra in 8 redshift bins at $3 < z_{\rm spec} < 10$. The red squares show the break strengths reported in \cite{RB24}, measured from 5 stacks at $5 < z_{\rm spec} < 10$. While $z < 4.5$ stacks exhibit strong Balmer breaks exceeding $f_{\nu}(4225{\rm\AA})/f_{\nu}(3565{\rm\AA})$ = 1.3, the break is almost entirely vanished by $z \sim 6$. Galaxies at $z > 7$ exhibit ``negative'' Balmer breaks, $f_{\nu}(4225{\rm\AA})/f_{\nu}(3565{\rm\AA}) < 1$, characteristic of very high specific SFRs and extremely young stellar populations. For comparison, the shaded regions show the expected break strengths for single stellar populations aged 1 Myr (blue), 10 Myr (yellow), and 100 Myr (red). The width of shaded regions indicates the range of the break strength as the metallicity varies from Z$_{\odot}$ to Z$_{\odot}$/10 at a fixed ionization parameter ($\log {\rm U} = -2$) and escape fraction ($f_{\rm esc} = 0$).}
    \label{fig: BB vs z}
\end{figure*}

\begin{figure*}
    \centering
    \includegraphics[width=13.5cm]{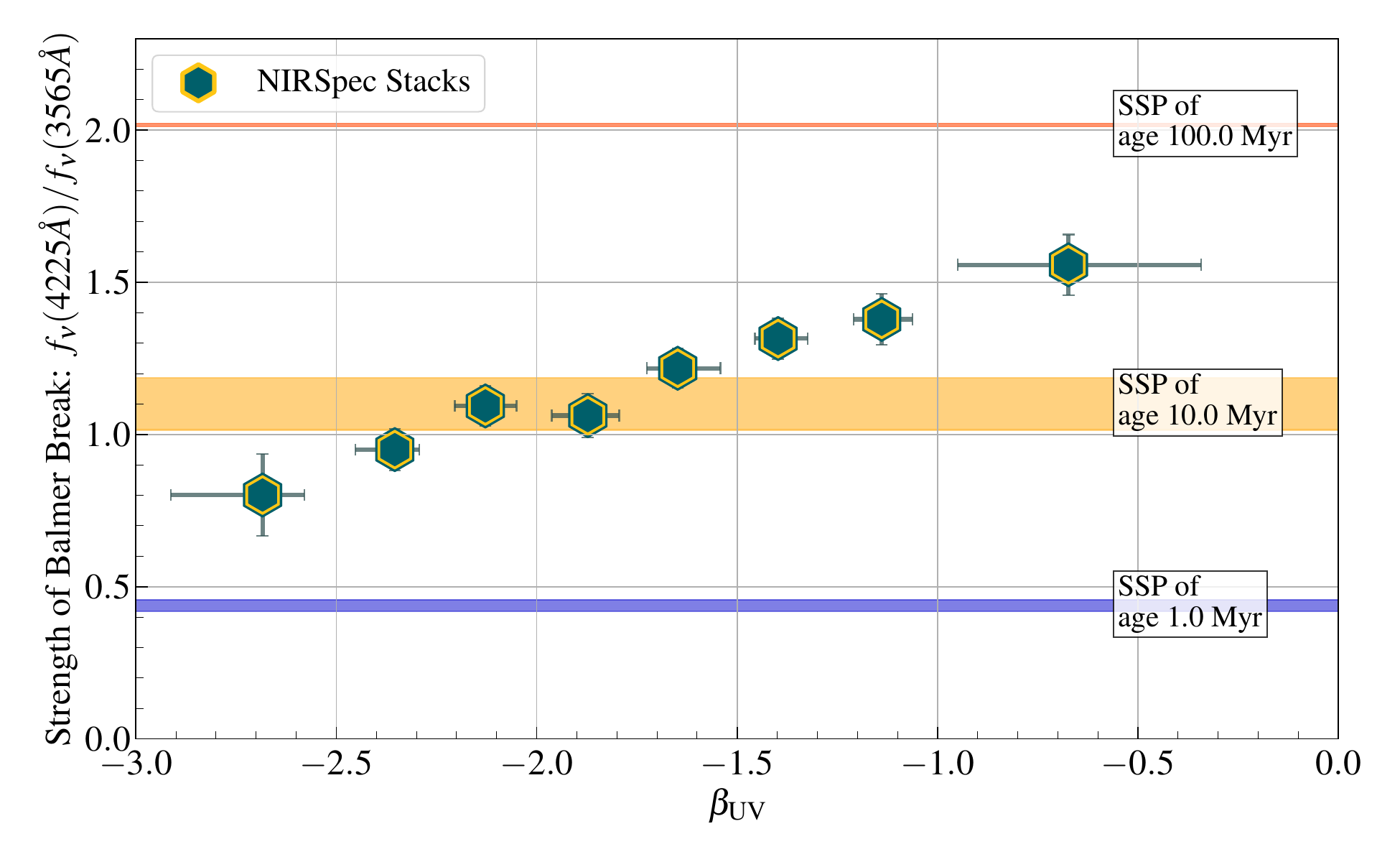}
    \caption{Correlation of the Balmer break strength and UV slope. The dark-green diamonds show the break strengths as measured from our stacks in 8 UV slope bins at $-3 < \beta_{\rm UV} < 0$. The stacks with $\beta_{\rm UV} > -1.75$, characteristic of relatively evolved and dusty stellar populations, show breaks exceeding $f_{\nu}(4225{\rm\AA})/f_{\nu}(3565{\rm\AA})$ = 1.2. The reddest stack with $-1 < \beta_{\rm UV} < 0$ exhibits the strongest Balmer break amongst both our redshift and UV slope stacks. The extremely blue stack with $\beta_{\rm UV} < -2.5$, characteristic of extremely young, metal-poor, and dust-free stellar populations with high escape fractions, shows a negative Balmer break with $f_{\nu}(4225{\rm\AA})/f_{\nu}(3565{\rm\AA}) < 1$.}
    \label{fig: BB vs beta}
\end{figure*}

The wavelength range of the common grid is chosen to inclusively cover the rest-frame spectra of all the galaxies in that bin. Since each bin comprises galaxies at different redshifts, the bluest and reddest wavelengths of the common grid are typically not covered in many objects; this is particularly pronounced in the $\beta_{\rm UV}$ bins which can include galaxies from both ends of our redshift coverage. In such scenarios, the bluest/reddest wavelengths of the stacks are only representative of the highest/lowest redshift galaxies in that bin, and not the entire sample in that bin. To avoid this, we only include those spectral pixels in our final stacks that are covered at least in $80\%$ of the galaxies in their corresponding bins.

\section{Balmer Breaks} \label{sec: Balmer breaks}

In this Section, we investigate the redshift and UV slope trends of the Balmer break strength ($\sim 3645$\AA) based on the stacked spectra. We define the break strength as the ratio of mean flux density ($f_{\nu}$) on the red side of the break to that on its blue side. The ``red'' and ``blue'' wavelength windows are chosen to avoid the emission lines; this choice is limited by the spectral resolution of the stacks, where spectral features are significantly broadened due to spectroscopic redshift inaccuracies. Similar to \cite{RB24}, we define the rest-$3565 \pm 35$\AA\ as the blue window and the rest-$4225 \pm 35$\AA\ as the red window. The measured Balmer break strengths, $f_{\nu}(4225{\rm\AA})/f_{\nu}(3565{\rm\AA})$, for the redshift and UV slope stacks are reported in Tables \ref{table: redshift stacks} and \ref{table: beta stacks}, respectively, and shown in Figures \ref{fig: BB vs z} and \ref{fig: BB vs beta}.

\subsection{Redshift Evolution of Balmer Break} \label{sub: redshift evolution}

Figure \ref{fig: BB vs z} shows the redshift evolution of Balmer breaks from $z = 10$ to $z = 3$. We measure negative breaks (i.e., $f_{\nu}(4225{\rm\AA})/f_{\nu}(3565{\rm\AA}) < 1$) for the two highest redshift stacks, representing galaxies at $6 < z < 10$. While the $6 < z < 7$ stack is consistent with a zero break (i.e., $f_{\nu}(4225{\rm\AA})/f_{\nu}(3565{\rm\AA}) = 1$) within $1\sigma$ uncertainties, the $7 < z < 10$ stack robustly exhibits a negative break. This is in agreement with the recent results of \cite{RB24}, where negative breaks are found in 4 redshift stacks at $z > 6$ (shown as the red squares in Figure \ref{fig: BB vs z}). The $5.5 < z < 6$ stack, representing the end of cosmic reionization, is consistent with a zero break. The remaining 5 stacks at $3 < z < 5.5$ exhibit positive Balmer breaks, monotonically becoming stronger from $z = 5.5$ to $z = 3$, where it reaches $f_{\nu}(4225{\rm\AA})/f_{\nu}(3565{\rm\AA}) \sim 1.5$. 

Although fully interpreting the observed Balmer breaks requires SED-fitting of the stacked spectra, comparing the inferred break strengths with expectations for single stellar populations (SSPs) can be informative. We focus on this comparison here while postponing the SED-fitting to Section \ref{sec: SED-fitting}, where we investigate the stellar population trends of the break strength.

\begin{figure}
    \centering
    \includegraphics[width=8.75cm]{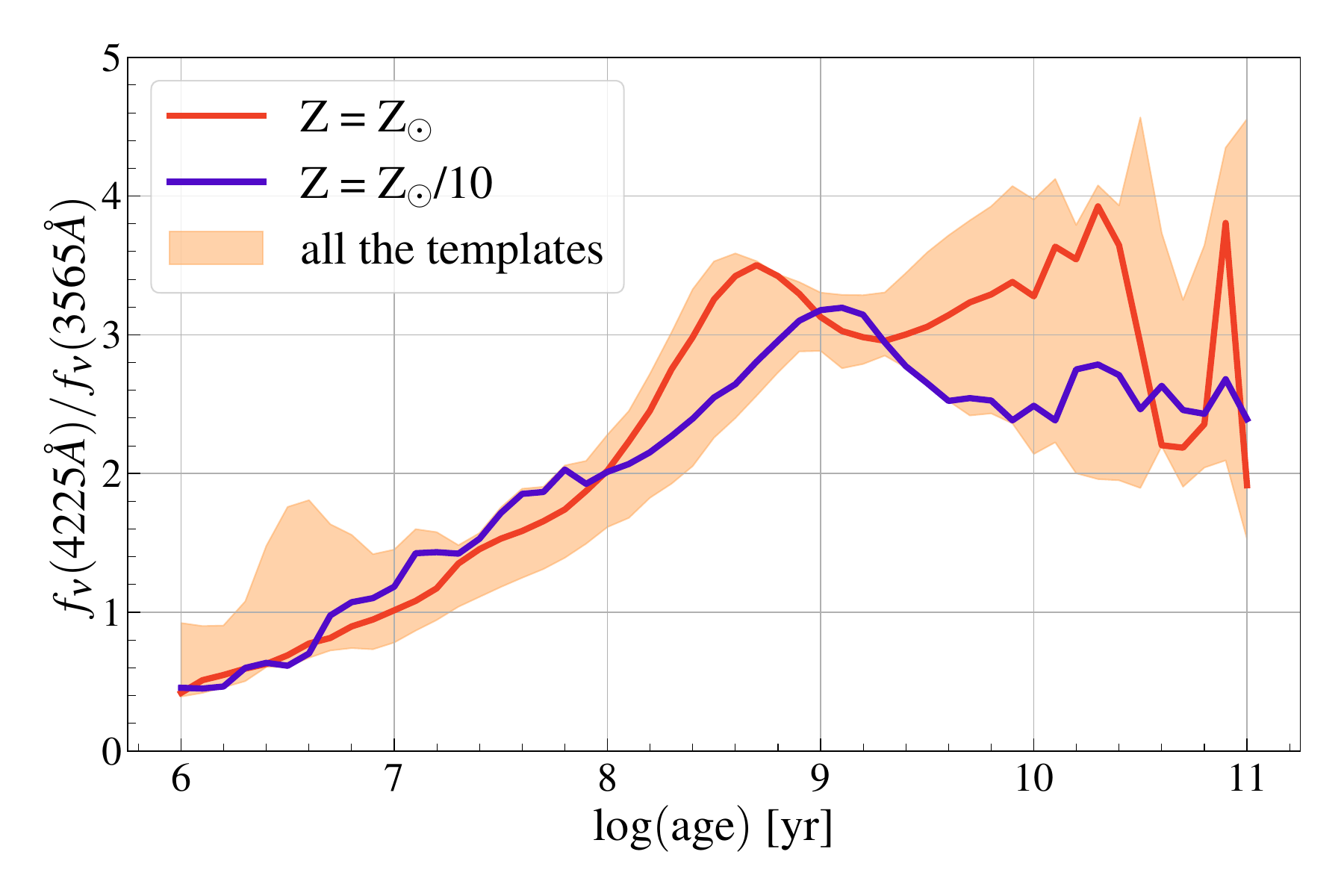}
    \caption{Age evolution of the Balmer/4000\AA\ break strength for single stellar populations (SSPs), inferred from a grid of stellar populations synthesis templates (see Section \ref{sub: redshift evolution}). The red and blue lines respectively show the break strengths for SSPs of ${\rm Z} = {\rm Z}_{\odot}$ and ${\rm Z} = {\rm Z}_{\odot}/10$ metallicity, at a fixed ionization parameter ($\log{\rm U} = -2$) and escape fraction ($f_{\rm esc} = 0$). The shaded region shows the variation in Balmer break strength for the covered range of metallicities, ionization parameters, and escape fractions (see Section \ref{sub: redshift evolution} for more details).}
    \label{fig: BPASS}
\end{figure}

We calculate the Balmer break strengths for SSPs in a range of ages (${\rm 1\; Myr} \leq {\rm age} \leq {\rm 100\; Gyr}$), stellar and gas-phase metallicities ($-5 \leq \log {\rm Z} \leq -1.4$), ionization parameters ($-4 \leq \log {\rm U} \leq -1$), and nebular escape fractions ($0 \leq f_{\rm esc} \leq 1$). For this purpose, we adopt the \textsc{bpass} stellar templates \citep{2017PASA...34...58E, 2018MNRAS.479...75S, 2022MNRAS.512.5329B} with a Chabrier IMF \citep{chabrier+2003}. We calculate the nebular emission for each template by running \textsc{cloudy} \citep{2023RMxAA..59..327C} over the aforementioned range of ionization parameters. For each age, metallicity, and ionization parameter, we model the non-zero escape fractions through linear combinations of the nebular and stellar-only templates. A more comprehensive description of these templates is provided in Langeroodi et al.\ (in prep.). We calculate the break strength for each template in the wavelength windows described earlier in this Section after convolving the spectra to the prism resolution.

Figure \ref{fig: BPASS} shows the age evolution of the Balmer/4000\AA\ break strength. The red and blue lines show the break strengths for ${\rm Z}_{\odot}$ and ${\rm Z}_{\odot}/10$ metallicity SSPs, respectively, both at a fixed $\log {\rm U} = -2$ and $f_{\rm esc} = 0$; the latter is more representative of $z > 6$ galaxies \citep{2023ApJ...957...39L, 2023ApJS..269...33N, 2023NatAs...7.1517H, 2023arXiv230408516C, 2024ApJ...962...24S, 2024arXiv240214084M}. The shaded region shows the variation in Balmer break strength for the covered range of metallicities, ionization parameters, and escape fractions. For SSPs, the break gets monotonically stronger up to $\sim 1$ Gyr, after which it plateaus around $f_{\nu}(4225{\rm\AA})/f_{\nu}(3565{\rm\AA}) \sim 3$. The shaded regions in Figures \ref{fig: BB vs z} and \ref{fig: BB vs beta} indicate the break strengths for SSPs of ages 1 Myr (blue), 10 Myr (yellow), and 100 Myr (red); the width of each region represents the change of break strength for a metallicity variation in the range ${\rm Z}_{\odot}/10 \leq Z \leq {\rm Z}_{\odot}$ (at a fixed $\log {\rm U} = -2$ and $f_{\rm esc} = 0$). 

We emphasize that the break strengths shown in Figure \ref{fig: BPASS} are only representative of ``single'' stellar populations, a scenario that is only descriptive of the simple galaxies at $z > 7$ where stellar populations are less likely to contain sizeable old components. Figure \ref{fig: BB vs z} shows that the measured break strengths for our two highest redshift stacks fall below that expected for a 10-Myr-old SSP. This indicates that these galaxies are dominated by extremely young stellar populations younger than 10 Myr. The lower-redshift stacks never exhibit Balmer breaks as strong as $f_{\nu}(4225{\rm\AA})/f_{\nu}(3565{\rm\AA}) = 2$, corresponding to a 100-Myr-old SSP. Although these galaxies have mass-weighted ages exceeding 100 Myr, the contribution of newly formed stars to their light results in shallower Balmer breaks. Fully understanding this interplay requires SED-modelling of these galaxies, which is the subject of Section \ref{sec: SED-fitting}. 

\subsection{Correlation of Balmer Break with UV Slope} \label{sub: slope evolution}

Figure \ref{fig: BB vs beta} shows the correlation of break strength with UV slopes. Balmer breaks get monotonically stronger as the galaxies become redder. Similar to the highest redshift stacks, the two bluest stacks with $-3 < \beta_{\rm UV} < -2.25$ exhibit negative Balmer breaks. Such extremely blue UV continua are characteristic of stellar populations with very high specific SFRs and potentially high escape fractions, where the light is dominated by unattenuated emission from young and metal-poor stars \citep[][Langeroodi et al.\ in prep.]{2023arXiv230708835T, 2023arXiv231106209C}. That the light in these galaxies is dominated by extremely young stellar populations is further confirmed by the inferred break strengths of these stacks, falling in the region between the 1-Myr- and 10-Myr-old SSP (blue and yellow shaded regions in Figure \ref{fig: BB vs beta}, respectively). 

The four stacks covering $-1.75 < \beta_{\rm UV} < 0$ exhibit positive Balmer breaks, exceeding $f_{\nu}(4225{\rm\AA})/f_{\nu}(3565{\rm\AA}) = 1.2$. The reddest stack with $-1 < \beta_{\rm UV} < 0$ exhibits the strongest Balmer break amongst both our redshift and UV slope stacks. Such red UV slopes are characteristic of relatively evolved, passive, and dusty galaxies, where because of the low specific SFRs the population of newly formed stars is not bright enough to outshine the older stellar population. However, as discussed in Section \ref{sub: redshift evolution}, even these red galaxies contain a sizeable population of younger stars resulting in Balmer breaks that are much shallower than expectations for SSPs of similar ages. Fully interpreting the ratio of young to old stellar components in these galaxies requires SED-fitting, which is presented in Section \ref{sec: SED-fitting}.

\section{SED Analysis} \label{sec: SED-fitting}

In this Section, we present the SED-fitting of the redshift and UV slope stacks to investigate the Balmer break strength trends with the inferred stellar population parameters. We describe our SED model in Section \ref{sub: SED} and the main results in Section \ref{sub: sSFR and Av}. 

\subsection{SED-fitting} \label{sub: SED}

We use the package from Langeroodi et al.\ (in prep.) to fit the stacked spectra. We adopt the \textsc{bpass} single stellar population (SSP) templates with a Chabrier IMF and add the nebular emission by running \textsc{cloudy} for a range of ionization parameters. Our \textsc{cloudy} calculations of emission line ratios, for a spherical-shell cloud geometry and a point-source of ionizing spectrum, are not guaranteed to reproduce the high-$z$ observations. Therefore, a subset of emission lines are decoupled from the \textsc{cloudy} nebular emission models and fitted directly to the spectra. These lines include Ly$\alpha$, [O\,\textsc{ii}]$\lambda\lambda$3727,29 doublet, H$\beta$, [O\,\textsc{iii}]$\lambda$4959, [O\,\textsc{iii}]$\lambda$5007, and H$\alpha$. After correction for attenuation, the measured H$\beta$ and H$\alpha$ flux are used to derive the SFRs where needed \citep[using the relation from][]{2013seg..book..419C}. 

We model the dust attenuation with a \cite{2000ApJ...533..682C} curve. We assume a similar level of stellar and nebular dust attenuation, $A_{\rm V,stars} = A_{\rm V,gas}$, to measure the unattenuated line fluxes. Observational evidence favours this choice for high-redshift star-forming galaxies over the local Universe calibration of $A_{\rm V,stars} = 0.44 \times A_{\rm V,gas}$ \citep[][and Langeroodi et al.\ in prep.]{2016ApJ...833..254S, 2024arXiv240116934B}. However, we note that the conversion factor might be smaller than 1 (i.e., $x < 1$ in $A_{\rm V,stars} = x \times A_{\rm V,gas}$). If so, our choice of $A_{\rm V,stars} = A_{\rm V,gas}$ would underestimate the unattenuated emission line fluxes and Balmer line SFRs. Therefore, our burstiness measure is conservative. 

We model the star-formation histories non-parametrically. Our last temporal bin spans $0-10^{6.5}$ yr in lookback time. The remaining bins are evenly spaced in 0.5 dex lookback time intervals from $10^{6.5}$ yr to the earliest onset of star formation, assumed to be $z = 20$. We calculate the maximum allowed lookback time by adopting the median redshift for each redshift stack (see Table \ref{table: redshift stacks}) and $z = 3$ for the UV slope stacks. We enforce a continuity star formation history prior, penalizing against sharp transitions in SFRs of immediate bins. For this purpose, we adopt the Student's-t prior from \cite{2019ApJ...876....3L}. 

The templates are convolved before fitting to the stacked spectra. The spectral resolution of the stacked spectra can significantly deviate from the pre-launch estimates for the NIRSpec prism. This is due to three reasons: i) The pre-launch line spread function (LSF) corresponds to a uniformly-illuminated slit. As shown in \cite{2023arXiv230809742D}, a more compact source morphology can result in spectral resolutions twice as high as the pre-launch measures. ii) Spectroscopic redshift measurement inaccuracies can broaden the spectral features in the stacks, resulting in a lower perceived spectral resolution. iii) Each stack consists of spectra over a wide redshift range. Therefore, each of its spectral bins has been probed with slightly different spectral resolutions in different galaxies (due to the wavelength-dependent spectral resolution of the NIRSpec prism). To account for these complexities, we treat the spectral resolution as a free parameter in our SED-fitting. To achieve this, we fit for a free parameter that inflates/deflates the pre-launch estimates of the prism resolution.

We explore the parameter space by employing the \textsc{dynesty} sampler \citep{dynesty, 2022zndo...6609296K}. \textsc{dynesty} adopts the dynamic nested sampling method developed by \cite{2019S&C....29..891H}. The best-fit parameters and uncertainties reported throughout this work correspond to the likelihood-weighted 16th, 50th, and 84th percentiles of the \textsc{dynesty} chains. The best-fit values are reported in Tables \ref{table: redshift stacks} and \ref{table: beta stacks}.

\subsection{Break Strength vs Specific SFR and Attenuation} \label{sub: sSFR and Av}

\begin{figure*}
    \centering
    \includegraphics[width=18.0cm]{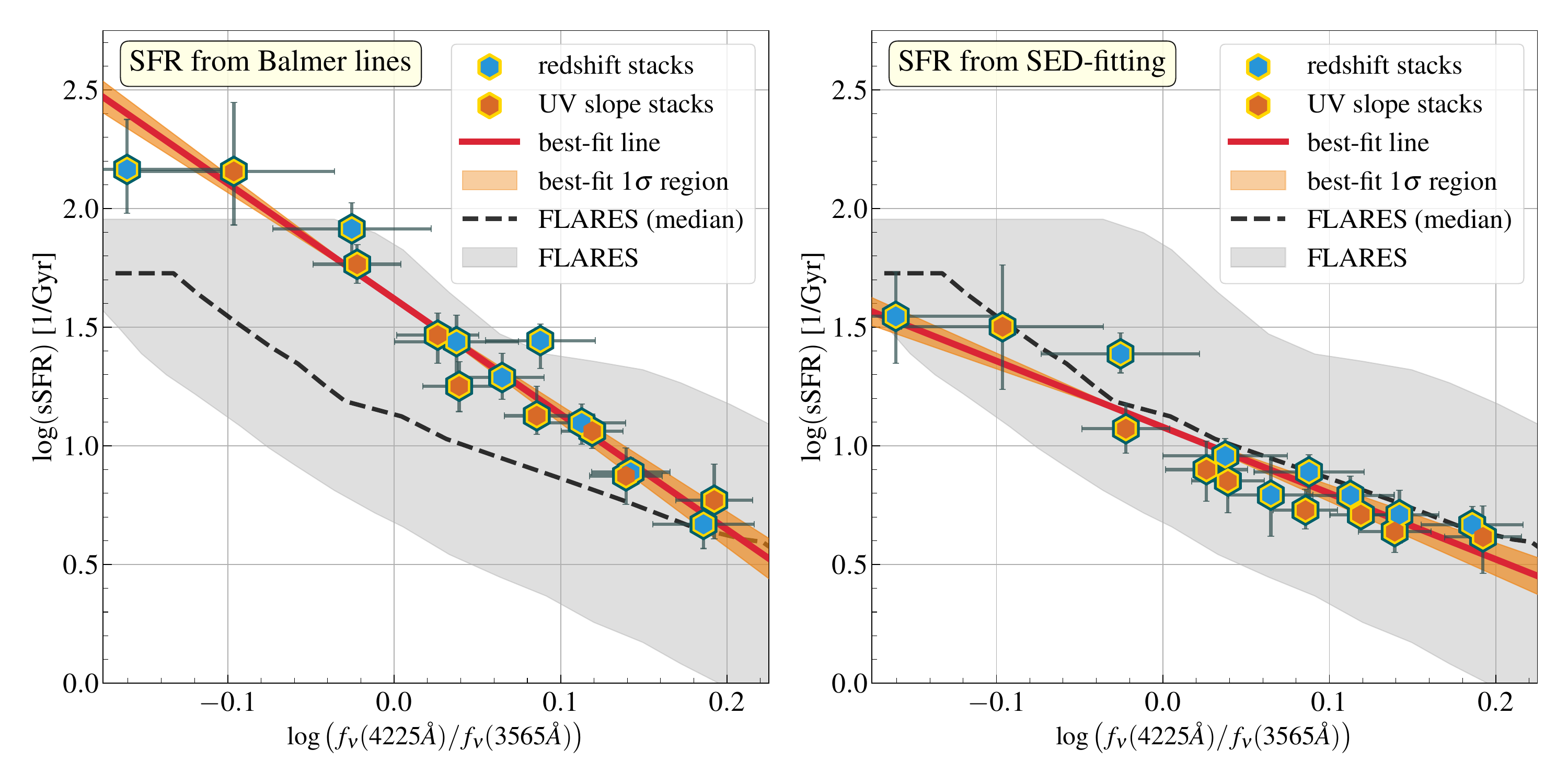}
    \caption{Anti-correlation between the sSFR and Balmer break strength. The blue and red diamonds show the measurements for our redshift and UV slope stacks, respectively. As shown, the sSFR is a strong predictor of Balmer break strength. In the left panel, SFRs are measured by fitting the Balmer lines. In the right panel, SFRs are measured through SED-fitting as the averages over the last 100 Myr of the best-fit non-parametric star formation histories. In each panel, the red solid line and the red shaded region show the best-fit linear relation and its $1\sigma$ uncertainty region, respectively. The best-fit relations are given in Equations \ref{eq: sSFR vs BB Balmer lines} and \ref{eq: sSFR vs BB SED-fitting}. The dashed black line and the grey shaded region show the predicted relation from the \textsc{flares} suite of hydrodynamical zoom-in simulations \citep{2024MNRAS.527.7965W}.}
    \label{fig: sSFR vs BB}
\end{figure*}

\begin{figure*}
    \centering
    \includegraphics[width=10.75cm]{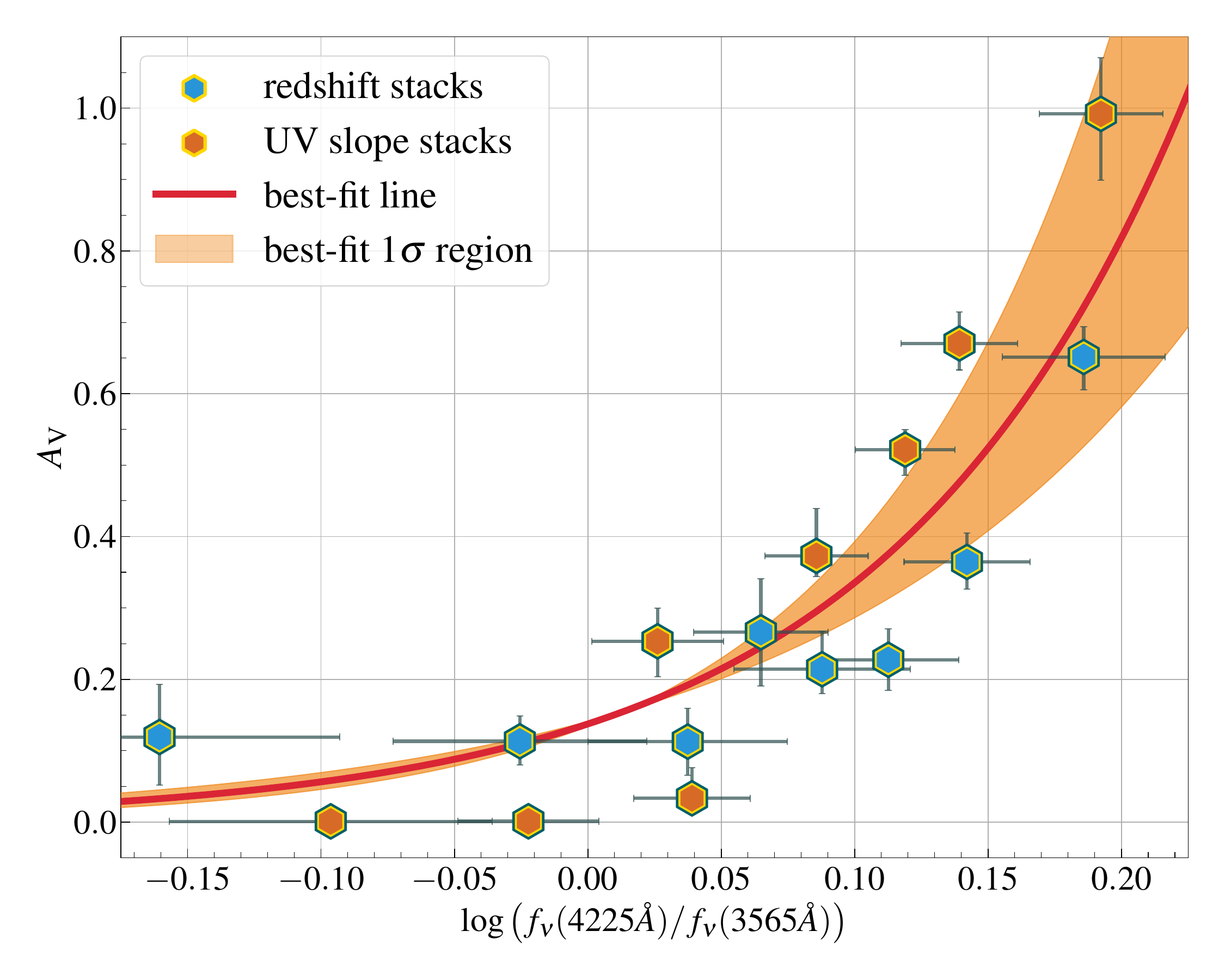}
    \caption{Correlation between the dust attenuation and Balmer break strength. The blue and red diamonds show the measurements for our redshift and UV slope stacks. For dusty galaxies, the attenuation is a good predictor of Balmer break strength. The solid red line and the red shaded region show the best-fit powerlaw and its $1\sigma$ uncertainty region, respectively. The best-fit relation is given in Equation \ref{eq: Av vs BB}.}
    \label{fig: Av vs BB}
\end{figure*}

\begin{figure*}
    \centering
    \includegraphics[width=12.25cm]{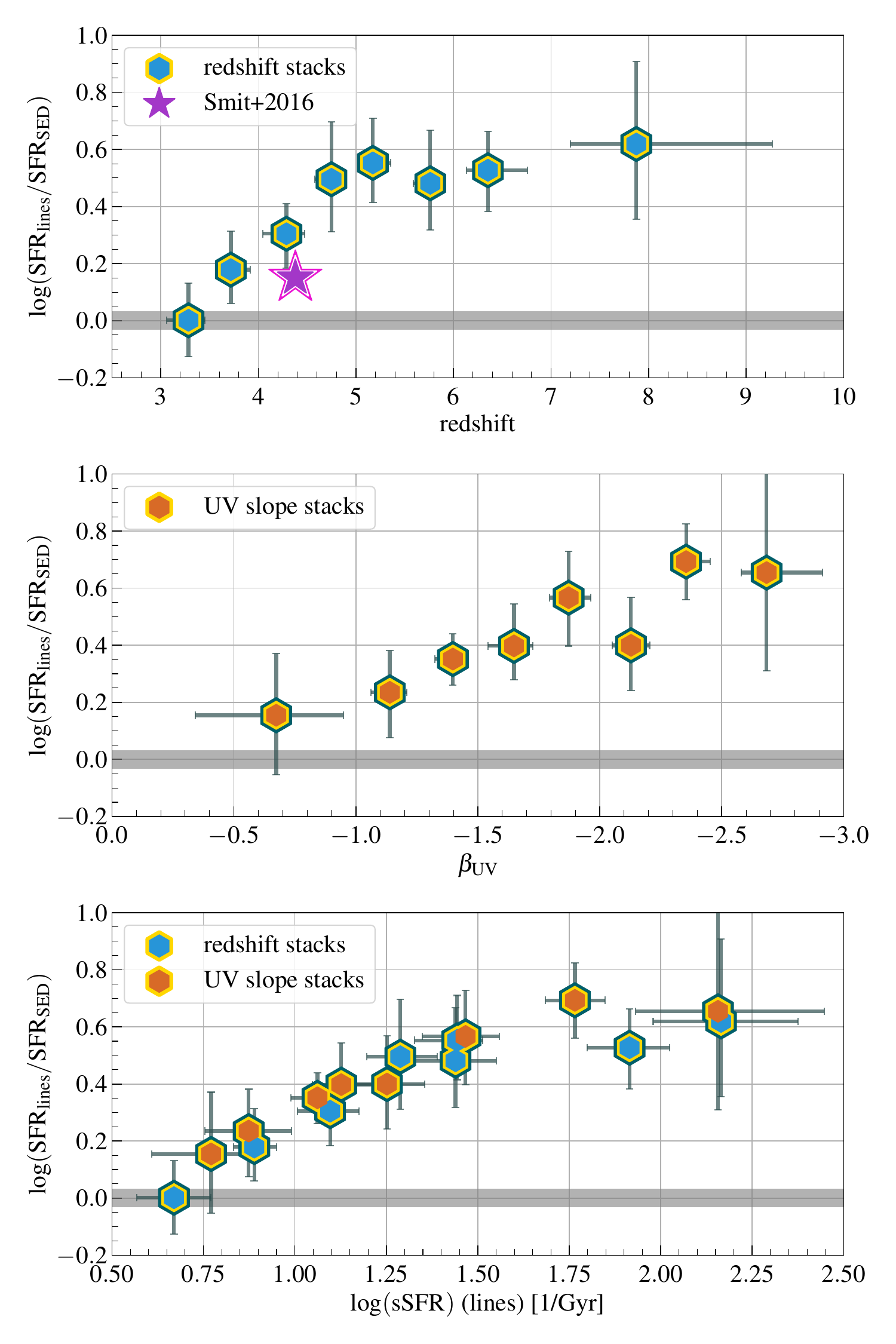}
    \caption{Transition from bursty to smooth star formation in high-redshift galaxies. We measure burstiness as the ratio of SFR over the past $\sim 10$ Myr to that over the past $\sim 100$ Myr; the former is measured from the Balmer lines (${\rm SFR}_{\rm lines}$) while the latter is measured through SED-fitting (${\rm SFR}_{\rm SED}$). A high ${\rm SFR}_{\rm lines}$ over ${\rm SFR}_{\rm SED}$ ratio indicates that galaxies undergo regular bursts of star formation on timescales shorter than a few 10 Myr. As shown in these panels, higher redshift galaxies, bluer galaxies, as well as those with higher specific SFRs exhibit more bursty star formation histories. Top panel: Blue diamonds show the measured SFR ratios for our stacked spectra in 8 redshift bins. The pink star shows a similar measurement from \cite{2016ApJ...833..254S} for a sample of galaxies at $z \sim 3.8-5$, with HST, Spitzer/IRAC, and ground-based observations. Middle panel: Orange diamonds show the measured SFR ratios for our stacked spectra in 8 UV slope bins. Bottom panel: Blue and orange diamonds show the measured SFR ratios for our redshift and UV slope stacks, respectively. The dominance of burstiness rises monotonically with redshift and specific SFR, while monotonically declining with UV slope. In each case, it plateaus at $\log({\rm SFR}_{\rm lines}/{\rm SFR}_{\rm SED}) \sim 0.6$.}
    \label{fig: SFR ratio}
\end{figure*}

\cite{2024MNRAS.527.7965W} explored the relationship between the Balmer break strength and key stellar population parameters based on the \textsc{flares} suite of zoom-in hydrodynamical simulations \citep{2021MNRAS.500.2127L, 2021MNRAS.501.3289V}. The explored parameters include specific star-formation rate (sSFR), dust attenuation, and age. \cite{2024MNRAS.527.7965W} showed that break strength is expected to be strongly and monotonically anti-correlated with sSFR, less strongly correlated with dust attenuation, and only weakly correlated with age. In particular, at $f_{\nu}(4225{\rm\AA})/f_{\nu}(3565{\rm\AA}) > 1.0$ the break strength is expected to be largely degenerate with age. Furthermore, contrary to stellar mass and dust attenuation, SED-fitting age inference is highly sensitive to the adopted SFH prior \citep[see e.g.][]{2019ApJ...876....3L}. Therefore, in this work, we mainly focus on the relation between break strength and specific star formation rate (sSFR), as well as dust attenuation.

Figure \ref{fig: sSFR vs BB} shows the anti-correlation between the break strength and sSFR. In the left panel, SFRs are measured from H$\alpha$ where covered, and H$\beta$ otherwise; both lines are fitted simultaneously while fitting the spectra (see Section \ref{sub: SED}). In the right panel, SFRs are measured through SED-fitting as the averages over the last 100 Myr of the best-fit non-parametric star formation histories. We find tight variance-weighted best-fit powerlaws in both cases, shown as the red lines and the red-shaded regions in Figure \ref{fig: sSFR vs BB}. For the Balmer line SFRs, we find 

\begin{align}
    & \log({\rm sSFR_{\rm lines}\; [Gyr}^{-1}]) = \notag \\
    & (-4.86 \pm 0.38) \log\bigg(\frac{f_{\nu}(4225{\rm\AA})}{f_{\nu}(3565{\rm\AA})}\bigg) + (1.26 \pm 0.03)\;; \label{eq: sSFR vs BB Balmer lines}
\end{align}

\noindent
while for the SED-fitting SFRs, we find

\begin{align}
    & \log({\rm sSFR_{\rm SED}\; [Gyr}^{-1}]) = \notag \\
    & (-2.79 \pm 0.34) \log\bigg(\frac{f_{\nu}(4225{\rm\AA})}{f_{\nu}(3565{\rm\AA})}\bigg) + (0.87 \pm 0.03)\;. \label{eq: sSFR vs BB SED-fitting}
\end{align}

\noindent
In general, the sSFRs measured from emission lines are systematically higher than those measured from SED-fitting. This is particularly the case for the stacks with the lowest break strengths, corresponding to those with the highest redshifts and bluest UV slopes (see Figures \ref{fig: BB vs z} and \ref{fig: BB vs beta}). This results in a higher normalization and a steeper best-fit relation for the Balmer line sSFR case. We discuss this further in Sections \ref{sec: burstiness} and \ref{sec: discussion}. 

The shaded grey region in Figure \ref{fig: sSFR vs BB} shows the expected anti-correlation between sSFR and break strength based on the \textsc{flares} simulations \citep{2024MNRAS.527.7965W}; the dashed black line shows the median \textsc{flares} relation. As shown in the right panel, our best-fit SED-fitting sSFR-vs-break strength relation closely aligns with the median \textsc{flares} relation. This is expected, since similar to the \textsc{flares}, we use the \textsc{bpass} SSPs. As shown in the left panel, while our best-fit Balmer line sSFR-vs-break strength relation agrees with the \textsc{flares} predictions, it gets increasingly offset from their median relation at lower break strengths. This corresponds to the highest redshift and blues rest-UV galaxies. We discuss this further in Sections \ref{sec: burstiness} and \ref{sec: discussion}. 

Figure \ref{fig: Av vs BB} shows the correlation between break strength and dust attenuation, where the latter is measured through SED-fitting (see Section \ref{sub: SED}). We find a variance-weighted best-fit powerlaw relation of the form 

\begin{align}
    & \log(A_{\rm V}) = \notag \\
    & (3.87 \pm 0.80) \log\bigg(\frac{f_{\nu}(4225{\rm\AA})}{f_{\nu}(3565{\rm\AA})}\bigg) + (-0.57 \pm 0.05)\;, \label{eq: Av vs BB}
\end{align}

\noindent
shown as the red line and the red-shaded region. Our stacks seem to probe the upper branch of the expected $A_{\rm V}$-vs-break strength correlation from the \textsc{flares} simulations \citep[Figure 9 in][]{2024MNRAS.527.7965W}. This branch corresponds to the intrinsically UV-bright galaxies, constituting the small fraction of dusty galaxies in \textsc{flares} simulations. We discuss this further in Section \ref{sec: discussion}.

We opted against measuring the dust attenuation through Balmer line ratios. This is because H$\alpha$ is covered only in 10 out of 16 stacks. Amongst these, the lowest redshift and the reddest stacks suffer from low spectral resolutions, insufficient to deblend H$\beta$ from [O\,\textsc{iii}]$\lambda\lambda$4959,5007 doublet. H$\gamma$ is not deblended from [O\,\textsc{iii}]$\lambda$4363 in any of our stacks, and the other Balmer lines are faint enough that their measured flux is highly sensitive to the best-fit spectral resolution. This results in only a few ($\sim$ 3 or 4) stacks where Balmer decrement can be robustly measured from emission lines. Moreover, the inference of stellar continuum attenuation from emission lines can be highly uncertain at high redshifts. As shown in \cite{2024arXiv240116934B} and Langeroodi et al.\ in prep., the high-$z$ conversion of the nebular emission line attenuation to the stellar continuum attenuation deviates from that measured for star-forming galaxies in the local Universe \citep[e.g.][]{2000ApJ...533..682C}. 

\section{Burstiness of Star Formation} \label{sec: burstiness}

In this Section, we evaluate the burstiness of star formation in high-redshift galaxies based on our redshift and UV slope stacks. We measure burstiness as the ratio of SFR measured from the Balmer emission lines (${\rm SFR}_{\rm lines}$) to that measured from the UV continuum through SED fitting (${\rm SFR}_{\rm SED}$). The former measures the star formation over the past $\sim 10$ Myr while the latter is defined (see Sections \ref{sub: SED} and \ref{sub: sSFR and Av}) to do so over the past $\sim 100$ Myr. A ${\rm SFR}_{\rm lines}$ to ${\rm SFR}_{\rm SED}$ ratio that is close to unity indicates that star formation has been relatively constant over a $\sim 100$ Myr timescale, corresponding to the smooth mode of star formation. However, a high ${\rm SFR}_{\rm lines}$ to ${\rm SFR}_{\rm SED}$ ratio indicates that galaxies undergo regular bursts of star formation on timescales shorter than a few tens of Myr, characteristic of the bursty mode of star formation. 

The top panel in Figure \ref{fig: SFR ratio} shows the redshift evolution of star formation burstiness. The blue diamonds show the measurements based on our 8 redshift stacks. The ${\rm SFR}_{\rm lines}$ to ${\rm SFR}_{\rm SED}$ ratio monotonically increases with redshift, indicating that bursty star formation becomes an increasingly more dominant mode of star formation for higher redshift galaxies. The $z > 4.5$ galaxies exhibit significantly bursty star formation histories, as evidenced by the $\sim +0.5$ dex offset of their ${\rm SFR}_{\rm lines}$ from their ${\rm SFR}_{\rm SED}$. The $z < 4.5$ galaxies do not exhibit particularly bursty star formation histories, as evidenced by their small ${\rm SFR}_{\rm lines}$ to ${\rm SFR}_{\rm SED}$ offsets. This is consistent with previous measurements by \cite{2016ApJ...833..254S} at $z = 3.8-5$, where a small $\sim +0.15$ dex ${\rm SFR}_{{\rm H}\alpha}$ to ${\rm SFR}_{\rm UV\;continuum}$ offset was interpreted as a slightly bursty mode of star formation at these redshifts. We note that the H$\alpha$ flux for a large portion of galaxies in \cite{2016ApJ...833..254S} was measured through SED-fitting of photometry-only data. However, \cite{2016ApJ...833..254S} confirmed their SED-fitting H$\alpha$ flux measurement technique on a calibration sample at lower redshifts (see Figure 13 in that study). Moreover, we note that similar to our work \cite{2016ApJ...833..254S} used a $A_{\rm V,stars} = A_{\rm V,gas}$ conversion to correct for dust attenuation.

The middle panel in Figure \ref{fig: SFR ratio} shows the UV slope trend of burstiness. The orange diamonds show the measurements for our 8 UV slope stacks. Bluer galaxies exhibit more bursty star formation histories. If the UV slope is taken as a specific SFR (sSFR) indicator (see Langeroodi et al.\ in prep.), based on this trend we expect a correlation between burstiness and sSFR. We test this in the bottom panel of Figure \ref{fig: SFR ratio}, which shows the correlation of burstiness with sSFR for our redshift and UV slope stacks. Bursty star formation becomes an increasingly more prominent mode for galaxies with higher sSFR. Since sSFR is strongly anti-correlated with stellar mass, this also indicates that bursty star formation becomes more common for lower stellar mass galaxies. In contrast, the smooth mode of star formation takes over in the redder, more mature galaxies with higher stellar mass. 

\section{Discussion} \label{sec: discussion}

\subsection{Appearance \& Evolution of Balmer Breaks} \label{sub: evolution of balmer breaks}

First, we investigate the appearance and redshift evolution of Balmer breaks and the implications of the observed break strengths for the corresponding stellar populations. Figures \ref{fig: BB vs z} and \ref{fig: BB vs beta} show the redshift and UV slope evolution of the break strength, respectively. The break gets monotonically stronger from $z = 10$ to $z = 3$, and from $\beta_{\rm UV} = -3.0$ to $\beta_{\rm UV} = 0.0$. 

We measure negative Balmer breaks (defined as $f_{\nu}(4225{\rm\AA})/f_{\nu}(3565{\rm\AA}) < 1$) at $z > 6$ and at $\beta_{\rm UV} < -2.25$. These stacks exhibit negligible dust attenuation (Figure \ref{fig: Av vs BB}). As indicated by the shaded regions in Figures \ref{fig: BB vs z} and \ref{fig: BB vs beta}, this picture is highly consistent with the expectations for extremely blue dust-free simple stellar populations that are younger than 10 Myr. This suggests that the star formation burst amplitude in these galaxies is high enough to completely outshine the older stellar population that might be in place \citep[e.g., see the discussion in][]{2024ApJ...961...73N}. This is confirmed by their extremely high sSFRs over the past $\sim 10$ Myr (Figure \ref{fig: sSFR vs BB}), sSFR $\gtrsim 30$ Gyr$^{-1}$, as inferred from the Balmer lines. In other words, the rest-UV to -optical light in these galaxies is completely dominated by star formation in the past $\sim 10$ Myr. 

Positive Balmer breaks start to appear in the stacked spectra at redshifts below $z = 5.5$. Similarly, we measure positive Balmer breaks at UV slopes redder than $\beta_{\rm UV} = -1.75$. These stacks exhibit more moderate sSFRs, none of which exceeds $\sim 20$ Gyr$^{-1}$ (Figure \ref{fig: sSFR vs BB}). Moreover, they seem to be more dust-obscured, with attenuation values exceeding $A_{\rm V} \sim 0.2$ mag and even as high as $A_{\rm V} \sim 1.0$ mag (Figure \ref{fig: Av vs BB}). This suggests that these stacks represent a more mature galaxy population, where the recent bursts of star formation are not extreme enough to completely outshine the older stellar population.

Amongst the stellar population parameters, we find the Balmer break to be strongly correlated with the sSFR, regardless of which method is used to measure the SFRs (i.e., Balmer lines vs.\ SED-fitting); this is shown in Figure \ref{fig: sSFR vs BB}. Following the above arguments, this suggests that the relative contribution of light from young stars to that from old stars is what determines the break strength. Moreover, we find that the break strength correlates strongly with the dust attenuation at $A_{\rm V} > 0.2$ mag (Figure \ref{fig: Av vs BB}). This is expected if the star-forming regions are enshrouded in more dust than the older regions; the relatively higher attenuation of young stars offsets their outshining effect and enables the rest-optical contribution of older stars to be seen, resulting in a stronger break.

We note that observational biases might influence the inferred trends in this work. Most of the galaxies in our sample are selected based on Ly$\alpha$ breaks in wide NIRCam imaging surveys before their NIRSpec prism follow-ups. Spectroscopic confirmation is biased toward galaxies with stronger UV continua and emission lines. Similarly, our variance-weighted stacking technique (Section \ref{sub: stacking}) might introduce biases toward galaxies with stronger UV continua. However, these effects are somewhat complex to estimate due to the diverse depths and observation strategies of NIRSpec follow-up campaigns, as well as magnification in the lensed fields. On aggregate, these can result in a bias toward galaxies with higher SFRs/sSFRs, especially at the higher redshift end where galaxies appear fainter in general. This can bias the measured Balmer break strengths low. For instance, this can explain the small offset between the measured break strength of our highest redshift stack and that in a similar redshift range from \cite{RB24} (see Figure \ref{fig: BB vs z}) where the spectra are weighted equally in producing the stacks. 

To further investigate the effect of observational biases, we revisited the break strength measurements of the UV slope stacks. We regenerated the UV slope stacks, this time by imposing redshift upper limits ranging from $z = 5$ to $z = 14$. We remeasured the break strengths for these stacks; the results are shown in Figure \ref{fig: obsbias}. The general trend between the break strength and UV slope remains intact. However, the two bluest stacks seem to be slightly biased toward lower break strengths for the stacks generated with higher redshift upper limits. We note that this trend can also be (partially) driven by the redshift evolution of the stellar populations, their ionization parameters, stellar and gas-phase metallicities, and nebular escape fractions, as shown by the shaded region in Figure \ref{fig: BPASS}. Nevertheless, the measured breaks for all the iterations agree within 1$\sigma$ uncertainties, and the offsets are much smaller than the dynamical range of Balmer breaks covered between the bluest and the reddest stacks. 

\subsection{Transition from Bursty to Smooth Star Formation} \label{sub: transition}

Here, we investigate the implications of our findings for the transition from bursty to smooth star formation. The three panels in Figure \ref{fig: SFR ratio} show that bursty star formation becomes increasingly more prominent at higher redshifts; for the bluer galaxies; and for those with higher sSFRs, which correspond to lower stellar masses. The ${\rm SFR}_{\rm lines}$ to ${\rm SFR}_{\rm SED}$ ratio seems to plateau at $\log({\rm SFR}_{\rm lines}/{\rm SFR}_{\rm SED}) \sim 0.6$, suggesting that we might be observing the peak of star formation burstiness at $z \gtrsim 6$, $\beta_{\rm UV} \lesssim -2.25$, and sSFR $\gtrsim 30$ Gyr$^{-1}$. 

Burstiness slowly declines from $z = 10$ to $z = 3$, as shown in the top panel of Figure \ref{fig: SFR ratio}. At redshifts below $z = 4$ galaxies do not appear particularly bursty, with $\log({\rm SFR}_{\rm lines}/{\rm SFR}_{\rm SED}) < 0.2$. This is in agreement with previous results from \cite{2016ApJ...833..254S}. It seems that for average galaxies the smooth mode of star formation starts taking over right before the cosmic noon. This transition from bursty to smooth star formation coincides with relatively redder galaxies and more moderate specific SFRs; $\beta_{\rm UV} > -1.25$ and ${\rm sSFR}_{\rm lines} \lesssim 10$ Gyr$^{-1}$. This corresponds to more mature and dust-obscured galaxies, with higher stellar masses. 

Similar to the ${\rm SFR}_{\rm lines}$/${\rm SFR}_{\rm SED}$ ratio, the break strength can also be interpreted as an indicator of star formation burstiness. This can be understood by thinking of bursty star formation as strong SFR temporal fluctuations around some mean value on timescales shorter than a few ten-Myr. In this context, the amplitudes of SFR fluctuations determine the degree of burstiness. In the most bursty galaxies with the strongest fluctuations, the bright $\sim 1-10$-Myr-old bursts with weak to negative breaks mostly outshine the older $> 100$-Myr-old stellar populations with stronger breaks (Figure \ref{fig: BPASS}). We identify these galaxies as the $z \gtrsim 4$ stacks, which exhibit break strengths consistent with a $1-20$-Myr-old stellar population. As mentioned above, these galaxies also exhibit high ${\rm SFR}_{\rm lines}$/${\rm SFR}_{\rm SED}$ ratios. The most bursty stacks exhibit negative Balmer breaks, characteristic of dust-free stellar populations that are younger than 10 Myr. On the contrary, in galaxies with smoother star formation histories corresponding to weaker SFR fluctuations, the light is not dominated by the most recent bursts. This allows the stronger Balmer break from $> 100$-Myr-old stellar populations to peek through. We identify these galaxies as the $z \lesssim 4$ stacks, with Balmer breaks stronger than $f_{\nu}(4225{\rm\AA})/f_{\nu}(3565{\rm\AA}) = 1.3$. 

Burstiness declines at lower sSFRs, as shown in the lower panel of Figure \ref{fig: SFR ratio}. Due to the strong anti-correlation between sSFR and stellar mass (e.g., see Langeroodi et al.\ in prep.), this suggests that galaxies with higher stellar mass exhibit smoother star formation histories. This indicates that as galaxies mature and build up their stellar mass, their ability to sustain star formation over longer timescales increases. Other lines of evidence support this conclusion. For instance, galaxies with smoother star formation histories exhibit redder rest-UV colors (middle panel of Figure \ref{fig: SFR ratio}), typically associated with more mature, dust-obscured stellar populations. This is in agreement with recent results by \cite{2023MNRAS.525.2241H}, based on a large suite of numerical simulations. These authors find that the increase in escape velocity promotes the smooth mode of star formation by preventing the escape of cool gas that is otherwise driven out by stellar feedback in smaller and less mature galaxies. 

Observational biases might influence our interpretations. As mentioned in Section \ref{sub: evolution of balmer breaks}, our spectroscopically confirmed sample is likely biased toward galaxies with stronger rest-UV continua. This can translate into a bias toward more bursty star formation histories. Our measure of burstiness is sensitive to the ratio of SFR over the past 10 Myr to that over the past 100 Myr. Based on our simple stellar population (SSP) templates (Section \ref{sub: SED}), a dust-free 1$-$10 Myr old SSP is roughly twice as UV-bright (at rest-2000\AA) as an 80$-$100 Myr old SSP of the same stellar mass. Therefore, as the galaxies become fainter with increasing redshift we expect to become increasingly more biased toward detecting more bursty galaxies. However, we argue that this does not significantly influence our findings, particularly because this trend is not seen in our redshift stacks. The top panel in Figure \ref{fig: SFR ratio} shows the redshift evolution of star formation burstiness. The redshift evolution mostly occurs between $z = 3$ and $z = 5$, while it plateaus at higher redshifts. If the observational biases were playing an important role in driving this redshift trend, we would expect a continued strong evolution at $z > 5$. 

\subsection{Implications for Photometry-Only Surveys} \label{sub: phot-only}

Here, we briefly discuss how our results based on a spectroscopically confirmed sample can be used to inform large imaging surveys. Constraining the star formation histories of galaxies and evaluating their burstiness requires accurate inference of their rest-UV flux/slopes, emission line equivalent widths, and Balmer break strengths \citep[see, e.g.,][]{2016ApJ...833..254S, 2023ApJ...952..143R, 2023arXiv231112691C}. These constrain the SFR over the past $\sim 100$ Myr, over the past $\sim 10$ Myr, and the stellar mass, respectively. The rest-UV properties are typically constrained accurately with photometry. However, the emission line equivalent widths and Balmer break strengths can be subject to degeneracies, particularly when only broad-band imaging is available. 

\begin{figure*}
    \centering
    \includegraphics[width=14cm]{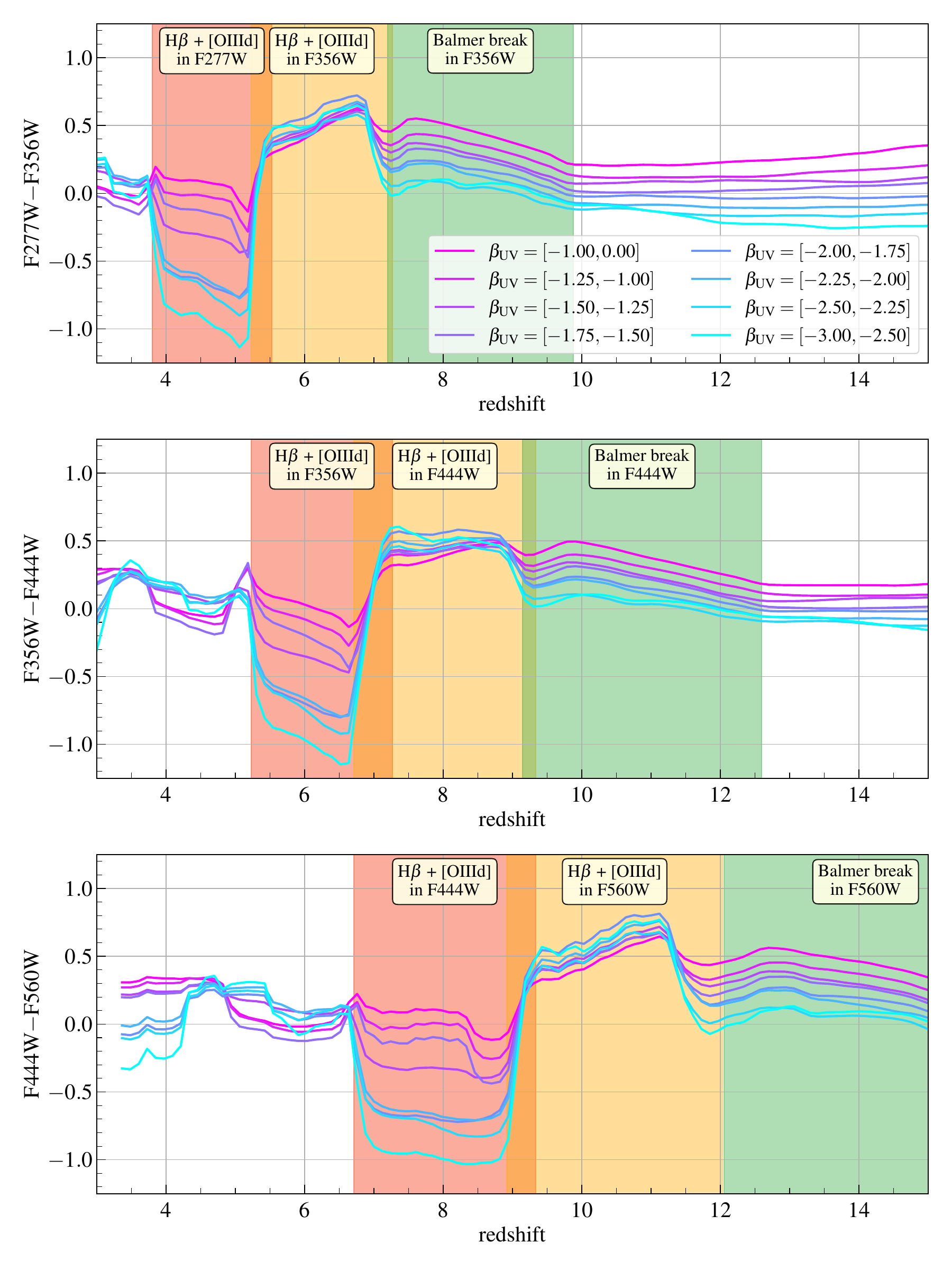}
    \caption{Redshift evolution of the NIRCam long-wavelength wide-band and MIRI F560W colors for our stacked spectra. The solid lines show the colors of our UV slope stacks, calculated by projecting their redshifted spectra onto the NIRCam and MIRI filters. The lightest-pink line indicates the colors of our UV-reddest stack with a break strength of $f_{\nu}(4225{\rm\AA})/f_{\nu}(3565{\rm\AA}) = 1.56$. The lightest-blue line indicates our UV-bluest stack with a break strength of $f_{\nu}(4225{\rm\AA})/f_{\nu}(3565{\rm\AA}) = 0.80$. The color-shaded regions mark the passage of prominent features through the filters. In the yellow region, the H$\beta$/[O\,\textsc{iii}]$\lambda\lambda$4959,5007 complex is captured by the red filter. In the green region, this feature is redshifted out of the red filter and instead replaced by the Balmer break. As shown in these panels, capturing the H$\beta$/[O\,\textsc{iii}]$\lambda\lambda$4959,5007 complex by the red filter has a similar but stronger effect on the colors than capturing the Balmer break. Therefore, accurate photometric redshifts are crucial for distinguishing between a Balmer break and strong emission lines. Moreover, as shown in Figure \ref{fig: BB vs beta}, the UV slope can be used as a proxy to predict the break strength and break this degeneracy.}
    \label{fig: color vs z}
\end{figure*}

Figure \ref{fig: color vs z} shows the expected near- to mid-infrared colors of our UV slope stacks placed at different redshifts. The colors are calculated by redshifting our stacked spectra and projecting them onto the NIRCam and MIRI filters. As shown, capturing the H$\beta$/[O\,\textsc{iii}]$\lambda\lambda$4959,5007 complex by the filters (the shaded yellow region in each panel) causes a similar color offset to that caused by capturing the Balmer break only (the shaded green region in each panel). This can make it challenging to distinguish between the bursty galaxies with strong emission lines and the more mature galaxies with smoother star formation histories and more pronounced Balmer breaks. Accurate photometric redshifts are key to breaking this degeneracy. The quality of photometric redshift estimation is subject to the photometric coverage, and especially the availability of medium-band imaging. However, medium-band imaging is not always available. 

\begin{figure*}
    \centering
    \includegraphics[width=16cm]{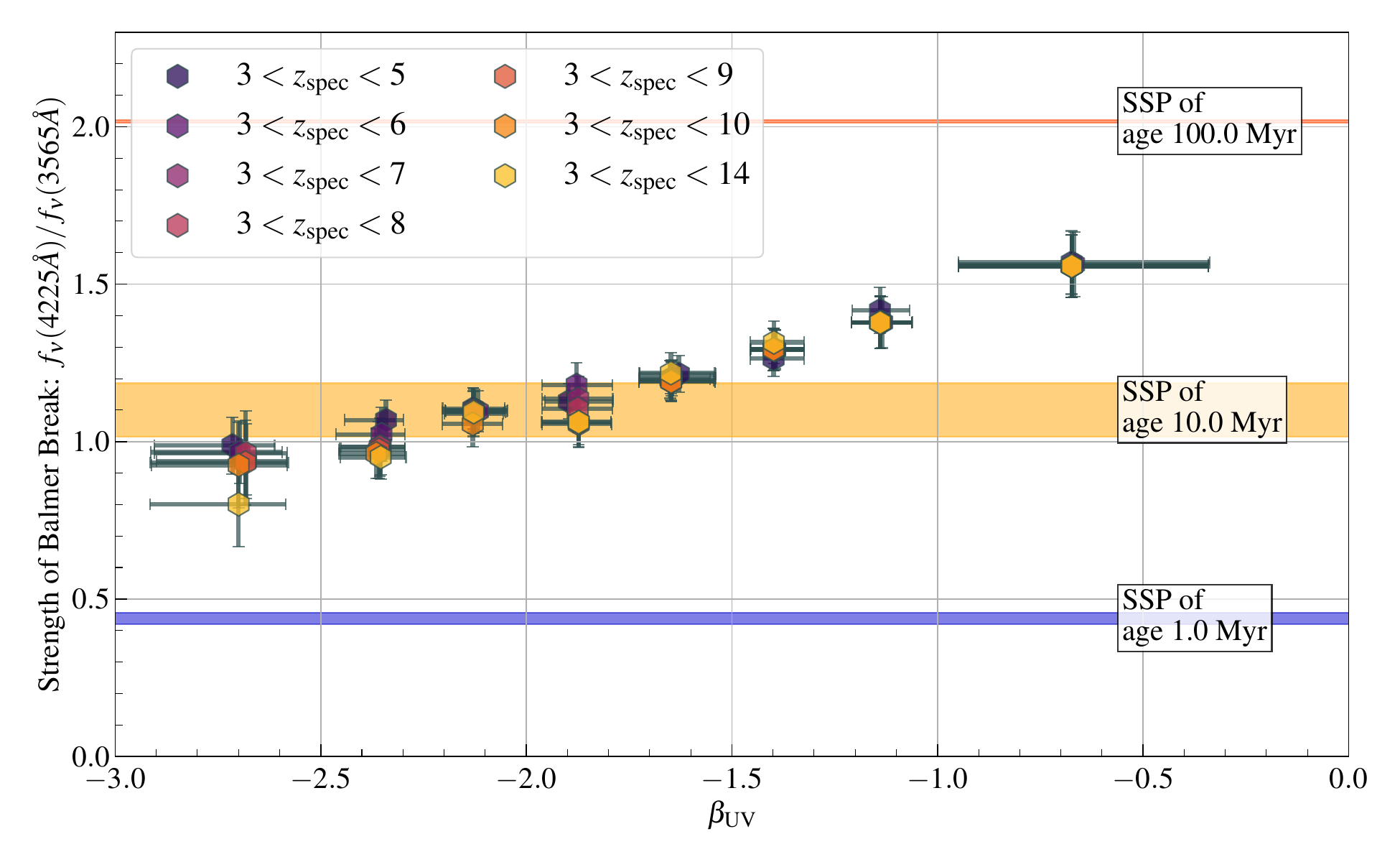}
    \caption{Same as Figure \ref{fig: BB vs beta}, but for regenerated UV slope stacks by imposing various redshift upper limits in the range $z_{\rm spec} = 5$ to $z_{\rm spec} = 14$. The measured Balmer break strengths are shown with the diamonds, color-coded with their upper limit redshift. The general trend of break strength vs.\ UV slope remains intact, while the two bluest stacks seem to be biased slightly toward lower break strengths for the higher redshift upper limits. Both the observational biases, as well as the redshift evolution of the stellar population, ionization parameter, stellar and gas-phase metallicities, and nebular escape fraction can contribute to this offset. For more details, see the discussion in Section \ref{sub: evolution of balmer breaks}.}
    \label{fig: obsbias}
\end{figure*}

Instead, as shown in Figure \ref{fig: BB vs beta}, the UV slope can be used as a proxy to accurately predict the break strengths. This can break or reduce their degeneracy with emission line equivalent widths. We note that, due to the redshift evolution of the UV slopes themselves, the UV slope trend seen in Figure \ref{fig: BB vs beta} can be partially driven by the redshift evolution of break strengths (Figure \ref{fig: BB vs z}). Nonetheless, based on Figure \ref{fig: obsbias} we argue that UV slope can be used as an accurate proxy for the break strength, regardless of the source redshift. This Figure shows that setting different upper limits on the redshifts of galaxies used for generating the stacks does not affect the inferred break strength vs.\ UV slope trend.

\section{Conclusion} \label{sec: conclusion}

In this work, we presented a uniform reduction and calibration of the publicly available NIRSpec prism spectra of 631 galaxies at $3 < z_{\rm spec} < 14$. We stacked these spectra in 8 redshift bins and 8 UV slope bins to investigate the redshift and UV slope trends of Balmer break strengths as well as star formation burstiness. We list our main conclusions below. 

i) The Balmer break strength monotonically increases from $z = 10$ to $z = 3$, and from $\beta_{\rm UV} = -3.0$ to $\beta_{\rm UV} = 0.0$. We find that the break strength is tightly anti-correlated with the specific SFR. This suggests that the relative contribution of light from young stars to that from old stars is what drives the break strength. We find that in dusty galaxies, with attenuation values above $A_{\rm V} = 0.2$ mag, the break strength is strongly correlated with the attenuation. This suggests that, in dusty galaxies, the systematically more dusty birthplace of young stars offsets their outshining effect and allows for the Balmer break from older stars to peek through. 

ii) We measure negative Balmer breaks ($f_{\nu}(4225{\rm\AA})/f_{\nu}(3565{\rm\AA}) < 1$) for the highest redshift as well as the bluest galaxies; i.e., at $z > 6$ and $\beta_{\rm UV} < -2.25$. These stacks exhibit negligible dust attenuation and extreme specific SFRs; ${\rm sSFR} > 30$ Gyr$^{-1}$. This picture is highly consistent with expectations for dust-free simple stellar populations that are younger than $\sim 10$ Myr. The rest-UV to -optical light in these galaxies is dominated by star formation in the past $\sim 10$ Myr. In contrast, positive Balmer breaks ($f_{\nu}(4225{\rm\AA})/f_{\nu}(3565{\rm\AA}) > 1$) appear at lower redshifts and for relatively redder galaxies; i.e., at $z < 5.5$ and $\beta_{\rm UV} > -1.75$. These stacks exhibit some dust attenuation and more moderate sSFRs; $A_{\rm V} > 0.2$ mag and ${\rm sSFR} < 20$ Gyr$^{-1}$. These stacks represent more mature galaxies, where the recent bursts of star formation are not extreme enough to completely outshine the older stellar population. 

iii) The bursty mode of star formation is more dominant at higher redshifts and lower stellar masses, corresponding to bluer galaxies and those with higher sSFRs. Our measure of burstiness seems to plateau at the highest redshift and lowest stellar mass end, suggesting that we might be observing the peak of star formation burstiness at $z \gtrsim 6$, $\beta_{\rm UV} \lesssim -2.25$, and sSFR $\gtrsim 30$ Gyr$^{-1}$. Burstiness slowly declines from $z = 10$ to $z = 3$. The $z < 4$ galaxies do not appear particularly bursty, suggesting that the smooth mode of star formation takes over right before the cosmic noon. The transition from bursty to smooth star formation coincides with relatively redder galaxies and more moderate sSFRs; i.e., $\beta_{\rm UV} > -1.25$ and ${\rm sSFR} < 10$ Gyr$^{-1}$.

iv) As galaxies mature, they transition from their bursty phase to a smooth phase of star formation. It seems that, as galaxies mature and grow in stellar mass their ability to sustain star formation over longer timescales increases. This is likely due to their increased ability to retain the cool gas that is otherwise ejected by star formation burst feedback in smaller and less mature galaxies. This evolution is supported by several lines of evidence. Our least bursty stacks exhibit the strongest Balmer breaks, the highest degree of dust attenuation, the reddest rest-UV colors, and the lowest sSFRs, all characteristic of relatively matured stellar populations. 

\software{
BPASS \citep{2017PASA...34...58E, 2018MNRAS.479...75S, 2022MNRAS.512.5329B}, 
cloudy \citep{2023RMxAA..59..327C}, 
corner \citep{2016JOSS....1...24F}, 
dynesty \citep{dynesty, 2022zndo...6609296K},
EAZY \citep{eazy},
msaexp \citep{msaexp},
scipy \citep{2020NatMe..17..261V},
SpectRes \citep{2017arXiv170505165C}
}

\section*{Acknowledgments}

D.L. greatly appreciates enlightening discussions with Mirko Curti and Viola Gelli, who inspired the investigation of bursty star formation histories. D.L. would like to extend gratitude to Pooneh Nazari who helped in preparing Figure \ref{fig: sSFR vs BB}. This work was supported by research grants (VIL16599, VIL54489) from VILLUM FONDEN.

\clearpage
\bibliography{main}
\bibliographystyle{aasjournal}

\end{document}